\documentclass[journal]{IEEEtran}
\usepackage{csquotes}
\usepackage{fancyvrb}
\usepackage[nolist]{acronym}
\usepackage{pifont}
\usepackage{graphicx}
\usepackage[export]{adjustbox}
\usepackage[super]{nth}
\usepackage{caption}
\usepackage{subfig}
\usepackage{float}
\usepackage{tabularx}
\usepackage{cite}
\usepackage{makecell}
\usepackage{array}
\usepackage{multirow}
\usepackage{booktabs}
\usepackage{amsmath}
\usepackage{threeparttable}
\usepackage[dvipsnames]{xcolor}
\usepackage[hyphens,spaces,obeyspaces]{url}
\usepackage[super]{nth}
\newcolumntype{C}{>{\centering\arraybackslash}X} 
\ifCLASSINFOpdf
\else
\fi

\begin{document}
\begin{acronym}
\acro {DNN}{deep neural network}
\acro {LRN}{local response normalization}
\acro {GAN}{generative adversarial network}
\acro {GNN}{graph neural network}
\acro {DLA}{deep learning accelerator}
\acro {PVA}{programmable vision accelerator}
\acro {VIC}{video image compositor}
\acro {SSD}{single shot detector}
\acro {GPC}{graphic processing clusters}
\acro {SM}{streaming multiprocessors}
\acro {DMA}{direct memory access}
\acro {VLIW}{very long instruction word}
\acro {VPU}{vector processing unit}
\acro {TPU}{tensor processing unit}
\acro {VPI}{Vision Programming Interface}
\acro {AxoNN}{energy-aware execution of neural networks}
\acro {HaX-CoNN}{heterogeneity aware execution of concurrent deep neural networks}
\acro {D-HaX-CoNN}{dynamic heterogeneity aware execution of concurrent deep neural networks}
\acro {CP-CNN}{computational parallelization for CNNs}
\acro {PCCS}{processor-centric contention-aware slowdown model}
\acro {PND}{partial network duplication}
\acro {SMT}{satisfiability modulo}
\acro {SAT}{satisfiability}
\acro {LP}{linear programming}
\acro {Jedi}{Jetson-aware embedded deep learning inference}
\acro {RNN}{recurrent neural network}
\acro {GA}{genetic algorithm}
\acro {CAN}{control area network}
\acro {H2H}{heterogeneous model to heterogeneous system mapping}
\acro {MaGNAS}{mapping-aware graph neural architecture search}
\acro {AI-SoC}{AI-System-on-Chip}
\acro {LLM}{large language models}
\acro {RL}{Reinforcement Learning}
\acro {PSNR}{peak-signal-to-noise ratio}
\acro {SSIM}{structural similarity ratio}
\acro {MSE}{mean-square error}
\acro {MRI}{magnetic resonance imaging}
\acro {CT}{computed tomography}
\acro {NPU}{neural processing unit}
\acro {FPGA}{field programmable gate array}
\acro {ASIC}{application specific integrated circuit}
\acro {DSP}{digital signal processor}
\acro {AI}{artificial intelligence}
\acro {PIDD-GAN}{parallel imaging coupled dual discriminator generative adversarial network}
\end{acronym}
\captionsetup[figure]{justification=centering}
%
\title{Edge GPU Aware Multiple AI Model Pipeline for Accelerated MRI Reconstruction and Analysis}

\author{Ashiyana~Abdul Majeed,~\IEEEmembership{Member,~IEEE,},
        Mahmoud~Meribout,~\IEEEmembership{Senior Member,~IEEE,}
        and~Safa~Mohammed Sali 
\thanks{Ashiyana Abdul Majeed, Dr Mahmoud Meribout, and Safa Mohammed Sali are with the Department
of Computer and Information Engineering, Khalifa University, Abu Dhabi, UAE (email: 100059454@ku.ac.ae, mahmoud.meribout@ku.ac.ae, safa.msali@ku.ac.ae).}}
\maketitle

\begin{abstract}
Heterogeneous edge devices integrate GPUs and domain-specific accelerators such as \ac{DLA} to meet the performance and energy demands of real-time \ac{AI} workloads. Many emerging applications, including medical imaging, industrial inspection, and autonomous systems, require multiple \ac{AI} models operating concurrently. However, existing optimizations focus either on single-model acceleration or on scheduling strategies that do not account for incompatibilities across heterogeneous accelerators. These incompatibilities frequently trigger fallback execution, reducing parallelism and overall throughput. This paper proposes a hardware-aware refinement methodology that substitutes accelerator-incompatible layers while preserving model fidelity, enabling complete execution on dedicated AI accelerators and eliminating performance degradation caused by fallback. The approach applies to convolutional architectures commonly used for image-to-image translation and object detection. To demonstrate effectiveness, we present a medical imaging case study involving CT-to-MRI image generation and concurrent diagnostic detection on NVIDIA's edge GPU Jetson Orin. By redesigning convolutional transpose operations to ensure DLA compatibility, the refined models eliminate GPU fallback and enable balanced multi-model scheduling. Experimental results show up to 170 frames per second in real-time execution of simultaneous diagnosis and MRI reconstruction (twice the performance compared to the original model) while maintaining structural similarity metrics at 74\%. These results validate that hardware-aware architectural refinement, when combined with heterogeneous scheduling, provides a scalable strategy for high-throughput edge AI pipelines.
\end{abstract}


%
\IEEEpeerreviewmaketitle

\section{Introduction}

\IEEEPARstart{T}{here} is growing interest in the design of heterogeneous, GPU-aware \acf{AI} models, where \ac{AI} architectures are explicitly tailored to the constraints and capabilities of embedded hardware platforms. Rather than deploying generic models and relying solely on runtime optimization, these approaches incorporate hardware constraints directly into the model design. For edge-based inference, \ac{AI} models must achieve low latency and high throughput while operating within strict power and resource budgets, all without compromising integration with upstream imaging systems. Modern edge platforms integrate specialized accelerators, including \acfp{NPU} and \acfp{DLA}, optimized for deep learning workloads. However, fully leveraging these heterogeneous resources requires careful alignment between the model architecture and the accelerator's capabilities. Efficient utilization of such accelerators is crucial to maintain high-throughput, low-latency, low-power processing pipelines, particularly in real-time imaging applications.

Multi-model pipelines are increasingly prevalent in real-world deployments, where multiple models are used concurrently within a single workflow to perform various tasks. For instance, \cite{retinal-imaging} integrates distinct models for classification, risk assessment, and logistic regression to support retinal diagnosis, while \cite{head-neck-cancer} combines an autoencoder, a vision transformer, and classifier networks for head and neck cancer staging. In medical imaging pipelines, common multi-model configurations include one network to perform image synthesis or reconstruction (e.g., CT-to-MRI translation), while another network simultaneously performs segmentation or abnormality detection. Such concurrent execution scenarios place substantial demands on heterogeneous edge platforms that incorporate multiple accelerators, including GPUs and \ac{DLA}. Devices such as the NVIDIA Jetson AGX Orin are designed to support parallel model execution across these compute units. However, achieving efficient multi-model execution requires careful coordination of hardware resources; otherwise, contention and suboptimal workload distribution can significantly degrade throughput and increase latency. 

\begin{figure*}[!t]
    \centering
    \includegraphics[width=1\linewidth]{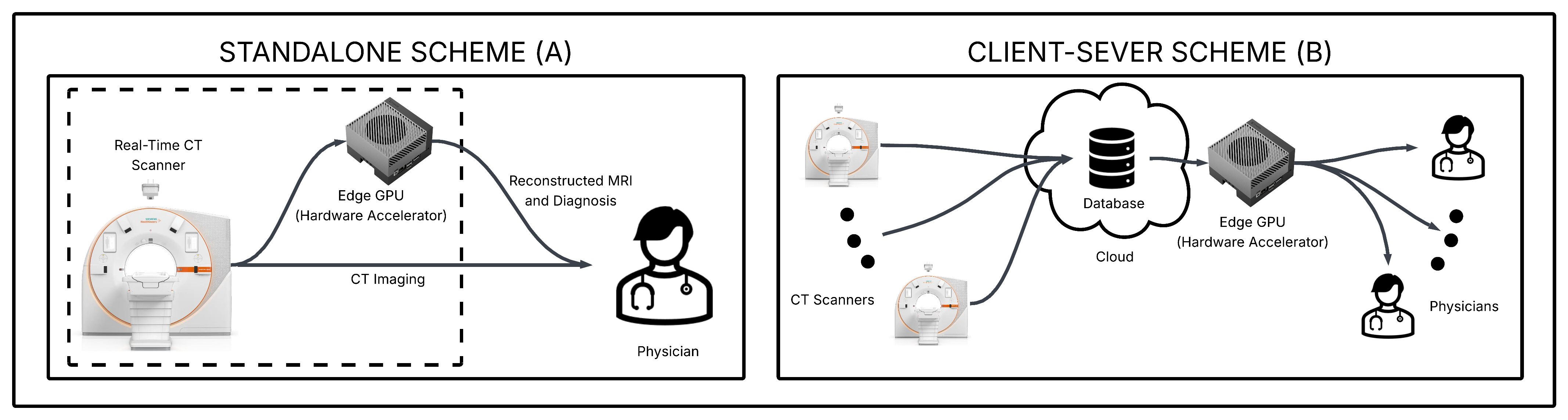}
    \caption{Proposed schema for the two methodologies}
    \label{fig:schemes}
\end{figure*}

To address these challenges, several schedulers have been developed. \ac{Jedi} \cite{Jedi} distributes model layers across GPUs and \acp{DLA} to improve utilization in real-time streaming applications. Similarly, \ac{HaX-CoNN} \cite{HaX-CoNN} extends this concept to concurrent deep neural networks, incorporating a \ac{PCCS}-based contention-aware model to minimize latency under shared-memory constraints. Nevertheless, the effectiveness of such schedulers in real-time scenarios remains application-dependent and requires further empirical validation. A GPU fallback can occur when running several models concurrently on a heterogeneous System-on-Chip device. Existing schedulers, including \ac{HaX-CoNN} and \ac{Jedi}, circumvent this issue by leveraging profiling data to pre-assign incompatible layers to the GPU. While this strategy prevents runtime fallback, it constrains the feasible transition points between \ac{DLA} and GPU execution, as all preceding layers must satisfy \ac{DLA} compatibility requirements. As a result, the search space for possible transition points is reduced, potentially constraining the optimization process and yielding an execution pipeline that is suboptimal. 

Operator incompatibility is a systemic issue across accelerators. Generally, \acp{NPU} and other accelerators require statically sized tensors and do not support dynamically sized tensors or models \cite{EyerissV2, TPU, Sima}. Google's \ac{TPU} and Edge \ac{TPU} do not support certain Python APIs, some of which are due to data type limitations, such as INT64 \cite{TPU}. As a result, the operations fall back on the CPU. The SiMa.ai platform imposes restrictions on layer parameters: transpose and concatenation operations cannot be executed on the batch axis. Pooling layers restrict the size to 128, and dilation to 1 \cite{Sima}. The issue of layer compatibility is, therefore, not confined to the \ac{DLA}. Consequently, even meticulously scheduled pipelines can exhibit degraded performance when GPU fallback events occur, as these events diminish effective parallelism across processing units \cite{11316238}. It is therefore essential to design AI-accelerator models with mechanisms that ensure full compilation and execution on the target device. These constraints highlight the need to account for both operator characteristics and tensor dimensionality when deploying \ac{DNN} models onto specialized acceleration hardware. While previous studies have explored lightweight model compression or quantization, fewer works explicitly address architectural refinement targeting accelerator compatibility.

The NVIDIA Jetson platform was selected as a representative hardware platform, as it encapsulates the defining characteristics of contemporary heterogeneous edge architectures, integrating a CPU, a programmable GPU, and dedicated AI accelerators within a single System-on-Chip. The same principle that was applied to the NVIDIA devices also applies to other comparable heterogeneous devices, such as the Libre \cite{Libre}. By focusing on a single, well-documented representative platform, we ensure controlled experimentation and fair model comparison under a consistent hardware baseline, while maintaining the generalizability of the proposed methodology to a broader class of edge AI devices.

This paper argues that hardware-aware architectural refinement is a prerequisite for efficient heterogeneous scheduling. Rather than treating fallback as a runtime artifact, we redesign incompatible layers to preserve model fidelity while enabling complete execution on dedicated accelerators. This strategy is architecture-agnostic and applicable to convolutional neural networks used in image-to-image translation, detection, and other vision tasks. To validate the general framework, we present two medical imaging case studies: CT-to-MRI generation alone and CT-to-MRI generation combined with real-time diagnostic detection. The medical domain is particularly suitable due to strict latency requirements and the prevalence of reconstruction analysis pipelines. However, the proposed methodology is not limited to this application and can be extended to broader edge AI systems that require concurrent multi-model execution. The main contributions of this paper are summarized as follows:
\begin{itemize}
    \item A hardware-aware reformulation of a CNN architectures that removes \ac{DLA}-incompatible operators, particularly padded transposed convolutions, while preserving structural fidelity and reconstruction accuracy. The proposed modification enables fully accelerator-compliant execution and prevents GPU fallback during multi-model scheduling.
    \item A detailed analysis of the architectural incompatibilities between standard CNN operators and \ac{DLA} execution constraints, identifying the possible causes of fallback behavior and quantifying their impact on accuracy, throughput and power efficiency.
    \item The design and validation of a medical imaging pipeline in which the modified model enables simultaneous \ac{MRI} reconstruction and diagnostic inference using \ac{CT} data. The modified model is also able to improve the parallel execution of \ac{MRI} reconstruction. The proposed approaches improve resource utilization and deterministic execution in multi-stream scenarios.
    \item A scheduling-aware deployment strategy that integrates model-level architectural adaptation with system-level pipelining and \ac{HaX-CoNN} scheduling principles \cite{HaX-CoNN}, demonstrating efficient concurrent execution of multiple CNN workloads on heterogeneous edge accelerators.
\end{itemize}

Fig. \ref{fig:schemes} summarizes the two different schemas for the hardware-accelerated pipeline. The standalone schema (Fig. \ref{fig:schemes}A) refers to real-time MRI reconstruction and diagnosis performed while CT imaging is ongoing. In this case, it is necessary to ensure that the reconstruction and diagnosis models run synchronously, with minimal delay between frames for both models. The client-server schema (Fig. \ref{fig:schemes}B) is an alternative that can be offered to medical centers. CT images are stored in a cloud database, where they can be processed by hardware accelerators for MRI reconstruction and diagnosis. Here, it does not matter if there is a delay between the models, as synchronizing is not necessary.

The remainder of this paper is structured as follows. Section II surveys related work on AI models and hardware accelerators. Section III presents the architecture of the NVIDIA Jetson platform and defines the performance metrics used for evaluation. Section IV describes the proposed edge GPU-aware implementation. Section V reports the accuracy results, and Section VI outlines the data collection methodology. Section VII presents the results for standalone execution. Sections VIII and IX detail the naïve and HaX-CoNN scheduling implementations, respectively. Finally, Section X discusses future research directions, and Section XI concludes the paper.

\section{Related Works}

\subsection{GAN-Based MRI Reconstruction Models and Diagnostic Models}
Various architectures have been designed for medical image transformation and diagnostic detection tasks. \ac{GAN}, diffusion, and transformer models are among the architectures explored \cite{GAN_vs_Diffusion,survey_MRI_CT}. This paper focuses on \ac{GAN} models, due to their fast inference and minimal memory requirements, while producing images of comparable quality \cite{accuracy_comparison}. Specifically, the Pix2Pix model can produce results comparable to those of the CycleGAN and U-Net models \cite{pix2pix-example}. Numerous studies have evaluated GAN-based transformation models using different types of datasets, including paired datasets (\ac{CT} scans with corresponding \ac{MRI} scans) and unpaired datasets (\ac{CT} scans combined with unrelated \ac{MRI} scans). Generally, models trained on rare paired datasets outperform those trained on unpaired datasets, and even adding a small number of paired images to an unpaired dataset can significantly enhance model performance \cite{pair-unpair,survey_MRI_CT}. \ac{GAN}s are also relatively more straightforward and faster to execute than architectures like diffusion models or transformers. For detection tasks, YOLO models have demonstrated strong performance \cite{YOLO_models}.

Although prior work has explored energy-efficient inference on heterogeneous SoCs and lightweight model design for embedded deployment, most studies—including those in medical imaging—optimize parameter count, pruning, or quantization without explicitly addressing operator-level accelerator compatibility. Moreover, throughput under concurrent workloads and energy-per-inference metrics are seldom reported for GAN-based medical reconstruction on edge devices. In particular, CNN-based medical reconstruction and detection models are typically ported to embedded GPUs without reformulating layers that violate DLA constraints (e.g., padded transposed convolutions, kernel-size limits, or tensor-shape restrictions). As a result, GPU fallback events occur during deployment, limiting achievable concurrency and degrading real-time throughput in multi-model pipelines. In contrast, this work targets architectural compatibility, ensuring that medical GAN-based reconstruction models execute fully on the DLA without fallback, thereby expanding the feasible scheduling design space for heterogeneous execution. This approach not only preserves reconstruction quality but also improves throughput and power efficiency on embedded platforms.

\subsection{Hardware-Aware CNN Reformulation for Accelerator Compatibility} 

Many models are not fine-tuned to run entirely on \ac{DLA} or other accelerators, causing unsupported layers to default to GPU execution, which introduces latency and undermines real-time performance. The idea of replacing \ac{DLA} incompatible layers was first suggested in \cite{layer_replacement}, where the adaptive average pooling layer in MobileNetV2 was replaced with average pooling. However, this does not cover issues resulting from the parameters rather than the layer itself, such as kernel size limitations. Rather than replacing the entire layer, partial modifications can be made to preserve the model's integrity and maintain performance comparable to the original. In addition to optimizing overall performance, layer replacement reduces the number of subgraphs generated by the engine plan. With concurrent execution of multiple models, the number of subgraphs may exceed the 16-subgraph limit \cite{tensorrt}, terminating execution. In contrast, the hardware-aware model mitigates this issue by reducing the number of generated subgraphs, enabling more efficient execution on the DLA.

While prior efforts such as EdgeYOLO \cite{edgeYOLO} and other lightweight detection models focus primarily on reducing parameter counts and computational complexity for edge deployment, they do not explicitly address operator-level compatibility with dedicated accelerators. The challenge of accelerator incompatibility is not limited to NVIDIA DLAs. 

Recent studies have demonstrated the benefits of accelerator-aware deployment for medical imaging workloads. \cite{Huang2022} proposed \ac{PIDD-GAN} for rapid MRI from sensitivity-weighted zero-filled MRI, achieving sub-5 ms inference times on the NVIDIA Titan RTX GPU and a SIMM of 94\%. The architecture employs a U-Net–style generator similar to the Pix2Pix model adopted in this work, but extends the adversarial training scheme by incorporating two discriminators. \cite{Endoscopy} demonstrated high-resolution MRI endoscopy using multi-GPU configurations, while \cite{tpu_vs_gpu} evaluated the energy efficiency of Edge TPU and embedded GPU platforms for ophthalmic imaging tasks. Although these works highlight the advantages of specialized hardware, they primarily focus on system-level acceleration rather than structural model adaptation. Reported metrics typically include inference time and reconstruction accuracy, while throughput under concurrent workloads and power consumption on embedded edge platforms receive limited attention. In contrast, this work emphasizes operator-level reformulation to eliminate incompatibilities at their source. By resolving fallback at the architectural level, the proposed methodology enables more effective scheduling and balanced heterogeneous execution.

\subsection{Multi-Model Scheduling Methods}

Many emerging heterogeneous inference frameworks for edge platforms implicitly assume that neural networks can be partitioned seamlessly across heterogeneous processors. In practice, however, scheduling decisions are often constrained by operator-level compatibility. Some approaches rely on profiling data to aggressively assign layers to accelerators such as the \ac{DLA}, even when partial incompatibilities exist, thereby restricting the adequate search space of feasible partitioning points \cite{my_review}. Therefore, beyond model compression or architectural scaling, there remains a need for systematic, hardware-aware model refinement that explicitly targets accelerator compatibility at the operator and parameter level. Our approach addresses this gap by ensuring that multi-model pipelines can execute without fallback transitions, thereby preserving concurrency and maximizing heterogeneous resource utilization.

\subsection{Hardware Accelerators in Medical Applications}
Capable edge devices are required for real-time processing in hospitals and clinics. Traditionally, CPU-based systems have been used in medical imaging. Although flexible, they face scalability limits and may be inefficient. In most cases, medical imaging systems use \ac{FPGA} to reconstruct \ac{MRI} and \ac{CT} images. However, GPUs are required when using more powerful AI models. Various other chip architectures, such as \ac{ASIC} and \ac{DSP}, have also been explored for other use cases, including ultrasound filtering and detectors \cite{hardware_real_time}. Depending on the algorithm type, a particular hardware configuration would be more appropriate. 

Recent studies have applied AI to edge devices for medical diagnosis and monitoring. For instance, \cite{Epileptic} developed an IoT-enabled EEG seizure detection framework using spike-statistical and spectral features with CNNs, achieving 98.48\% accuracy on resource-constrained platforms. Similarly, \cite{Polyp} proposed Explainable Multitask Shapley Explanation Networks (EMSEN) for real-time polyp detection in colonoscopy videos, combining channel attention with Shapley-based interpretability to improve accuracy and interpretability. Both these works underscore the importance of combining domain-specific AI models with hardware-aware optimization strategies. These works follow a sequential methodology rather than executing multiple models in parallel.

\subsection{AI Models in Various Applications} 

Similar heterogeneous hardware trends are observed beyond medical imaging. In autonomous driving, robotics, industrial inspection, and intelligent surveillance systems, embedded platforms increasingly integrate CPUs, GPUs, NPUs, and domain-specific accelerators within a shared memory System-on-Chip. These systems must simultaneously execute perception, reconstruction, detection, and control tasks under strict latency and energy constraints \cite{MTPerception_AD,computing_acceleration}. \cite{es1865} reviews real-time object detection algorithms for autonomous vehicles and associated hardware accelerators, highlighting that modern convolutional neural networks are deployed on GPUs and ASICs to achieve maximum throughput. However, the paper also notes that achieving real-time performance across all sensor streams in a production autonomous system requires additional hardware and algorithmic innovation beyond single-accelerator designs.

In medical imaging, algorithmic diversity implies that no single accelerator is universally optimal. Classical signal-processing tasks often benefit from FPGA-style spatial parallelism, while convolutional neural networks achieve higher throughput on NPUs or GPUs optimized for dense matrix operations \cite{FlexNPU,EyerissV2}. When evaluating power efficiency, the \ac{FPGA} and \ac{NPU} consume less energy than the CPU and GPU. In particular, the \ac{NPU} can achieve low energy consumption with the ResNet50 model. In contrast, the \ac{FPGA} is well-suited for the other pre-processing and image processing tasks due to its architecture \cite{computing_acceleration}. Energy–latency trade-offs are similarly reported in other edge-AI deployments, where specialized accelerators such as NPUs or TPUs achieve significantly lower energy per inference than general-purpose GPUs for convolutional workloads \cite{tpu_vs_gpu}. However, naively assigning workloads to accelerators without considering inter-processor transitions, shared memory contention, and operator compatibility may result in suboptimal global performance, even when individual accelerators are energy-efficient. This phenomenon is particularly critical in multi-model pipelines where heterogeneous processors execute concurrently and compete for shared memory bandwidth \cite{HaX-CoNN}. Scheduling techniques such as Jedi \cite{Jedi} and HaX-CoNN \cite{HaX-CoNN} can address these issues and further optimize execution.

\subsection{Limitations of Existing Research}

There is limited focus on optimizing the execution of multiple models across heterogeneous edge devices in the medical imaging sector. Scheduling techniques have been tested only for \ac{DNN} models, not for any specific application. This paper addresses these gaps by proposing an edge GPU-aware multi-model pipeline for simultaneous MRI reconstruction from CT images and real-time diagnostic analysis. Unlike prior medical edge deployments, the proposed framework eliminates architectural-level layer incompatibilities, thereby reducing fallback transitions, increasing effective throughput under concurrent execution, and improving power efficiency on heterogeneous embedded platforms.

\section{Background}

\subsection{Hardware Architecture}

\begin{figure}[H]
    \centering
    \includegraphics[width=0.95\linewidth]{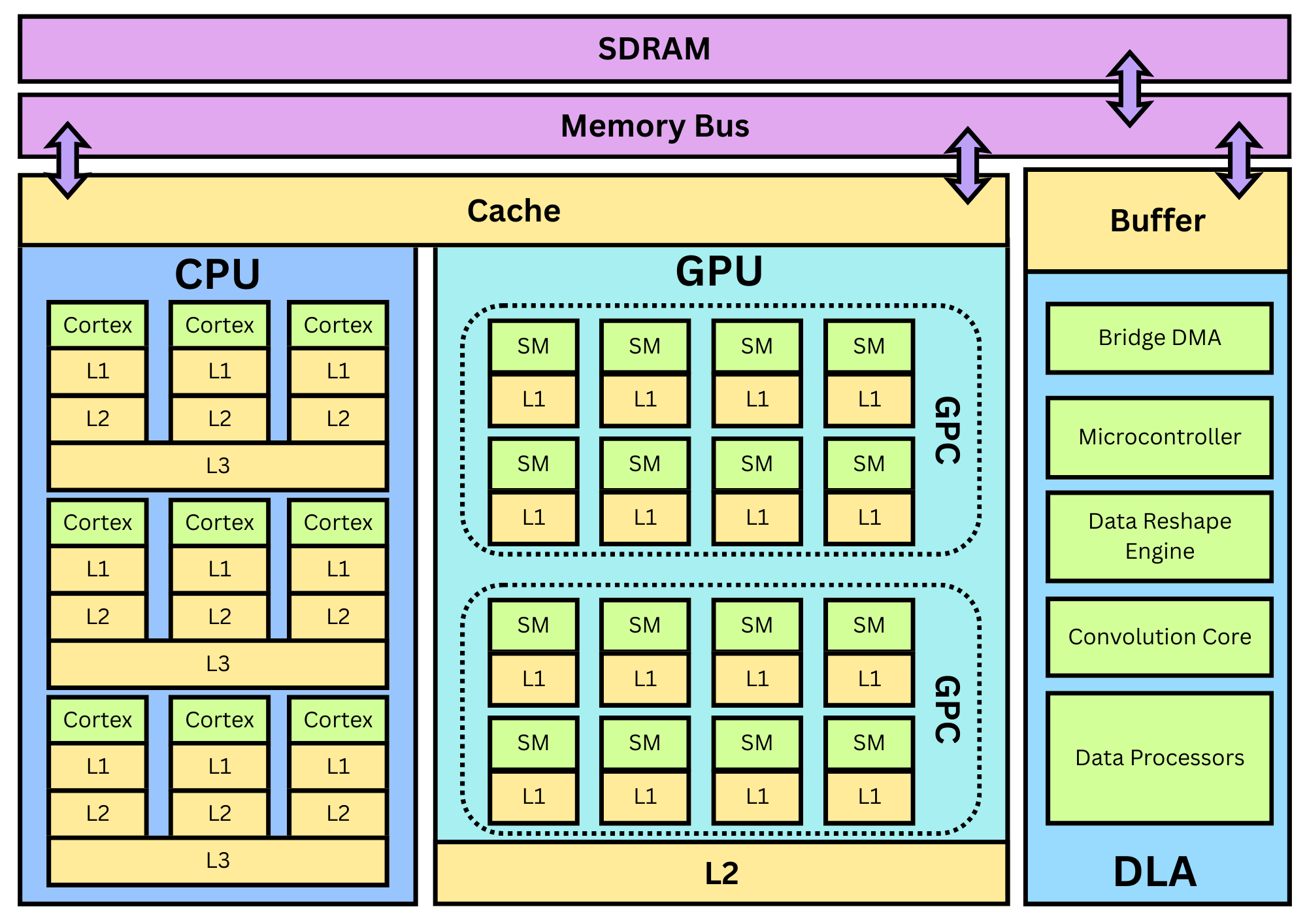}
    \caption{Block diagram of the NVIDIA Jetson AGX Orin}
    \label{fig:hardware}
\end{figure}

The NVIDIA AGX Orin achieves 1908 inferences per second with the Dashcam Net model \cite{xavier-orin-data}. The device combines multi-core ARM CPUs, an Ampere-based GPU, dual \ac{DLA} engines, and additional multimedia accelerators such as the \ac{PVA} and \ac{VIC}, all interconnected through a shared memory subsystem. Fig. \ref{fig:hardware} represents a block diagram of the hardware in the Jetson devices.

\subsubsection{\textbf{GPU}}
The Orin device uses an Ampere-architecture GPU designed for high-throughput parallel computation. The Ampere GPU consists of 2 \acp{GPC}, with each \ac{GPC} having 8 \acp{SM}. Each \ac{SM} contains 192 KB of L1 cache, 128 CUDA cores, and 64 Tensor cores. Tensor Cores within each SM accelerate mixed-precision matrix multiply-accumulate operations. Compared to standard CUDA cores, Tensor Cores provide significantly higher throughput per watt for structured linear algebra operations common in CNNs and transformers. Logically, the \ac{SM} uses a Harvard-style with separate instruction and data cache. The L2 cache is larger, with a size of 4 MB, enabling improved data locality for deep neural network workloads. It is connected to the memory using the memory subsystem interface. The GPU module alone can yield up to 170 Sparse TOPS to run AI models. \cite{xavier-orin-data}.  While highly flexible, it incurs higher energy consumption compared to fixed-function accelerators and may suffer from under utilization when executing small or highly structured convolutional workloads. 

\subsubsection{\textbf{\ac{DLA}}}
The \ac{DLA} is a fixed-function convolutional accelerator optimized for energy-efficient inference of CNN-based workloads. It comprises a microcontroller, a convolution core, data processors, dedicated memory, and data reshape engines \cite{nvdla-doc}. Unlike the GPU, which executes general CUDA kernels, the \ac{DLA} processes supported neural network operators through a hardware pipeline tailored for convolution, activation, and pooling layers. The 1 MB local buffer improves performance by a factor of 9 compared to the Xavier device. Since each accelerator is designed to perform specific functions, it has limitations when executing the different layers of an AI model. According to \cite{dla-limit}, when looking at the DLA in particular, some of the constraints include but are not limited to: 
\begin{itemize}
    \item Only FP16 and INT8 precision modes are supported. For equal operation, only INT8 is supported. For Slice and SoftMax operations, only FP16 is supported. 
    \item For deconvolution layers, padding must be zero. 
    \item Kernel sizes must range from 1 to 32. 
    \item Only two spatial dimensions are supported and the total number of dimensions including batch and channel is four.  
    \item The number of channels cannot exceed 8192. 
    \item The following activation functions are supported: ReLU, Sigmoid, TanH, Clipped ReLU, and Leaky ReLU. 
    \item Concatenation is only supported on the batch axis. 
    \item The \ac{LRN} layer only allows window sizes of 3, 5, 7, 9.
    \item For the unary layer, only ABS operation is allowed. 
    \item For the element-wise operations, sum, sub, product, max, min, and equal are supported. 
\end{itemize}

These constraints imply that networks containing unsupported operators or parameter combinations trigger fallback execution on the GPU. In multi-model pipelines, such fallback events disrupt concurrency, increase memory contention, and reduce effective throughput. Therefore, understanding and adapting to \ac{DLA} architectural limitations is essential for achieving balanced heterogeneous execution on Orin.

\subsection{Performance Metrics}
The pipeline's performance is assessed using throughput and latency. Throughput is the number of image frames the AI model can process per unit time, while latency is the time required to process a single image. In the image reconstruction case, accuracy is assessed using \ac{PSNR}, \ac{SSIM}, and \ac{MSE}. The \ac{MSE} is the cumulative squared error between the actual and reconstructed images. 
\begin{align} \label{eq:mse}
    MSE = \frac{1}{mn}\sum_{i=0}^{m-1}\sum_{j=0}^{n-1}\left ( O(i, j) - G(i, j)\right )^{2}
\end{align}
\begin{itemize}
    \item $m$ represents the height of the image.
    \item $n$ represents the width of the image.
    \item $O(i, j)$ represents the pixel value of the original image at the position $(i, j)$.   
    \item $G(i, j)$ represents the pixel value of the generated image at the position $(i, j)$.   
\end{itemize}
The \ac{PSNR} measures the ratio between the peak fluctuation and the \ac{MSE}. 
\begin{align} \label{eq:psnr}
    PSNR = 10 \cdot \log_{10}\Big(\frac{(L-1)^2}{MSE}\Big)
\end{align}
\begin{itemize}
    \item $L$ is the maximum intensity level of the image. 
\end{itemize}
\ac{SSIM} is a measure that compares images by measuring their luminance, contrast, and structural information. 
\begin{align} \label{eq:ssim}
    SSIM = \frac{(2\mu_O\mu_G + C_1)(2\sigma_{OG} + C_2)}{(\mu_O^2 + \mu_G^2 + C_1)(\sigma_O^2 + \sigma_G^2 + C_2)} 
\end{align}
\begin{itemize}
    \item $\mu_O$ and $\mu_G$ are the average luminance of the original and generated images respectively. 
    \item $\sigma_O^2$ and $\sigma_G^2$ are the variances of the original and generated images respectively.
    \item $\sigma_{OG}$ is the covariance between the two images.
    \item $C_1$ and $C_2$ are constants. 
\end{itemize}

\section{Proposed Methodology}

\subsection{GAN Model}

\subsubsection{\textbf{Model Architecture}}
\begin{figure}[H]
    \centering
    \includegraphics[width=0.7\linewidth]{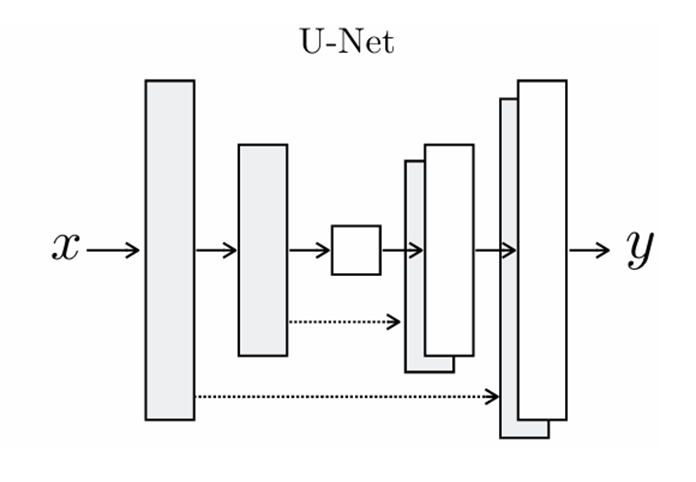}
    \caption{Simplified architecture of the GAN Pix2Pix model \cite{pix2pix}}
    \label{fig:Pix2Pix}
\end{figure}

The generative model considered in this study is the Pix2Pix conditional \ac{GAN}, composed of a U-Net generator and a PatchGAN discriminator \cite{pix2pix}. Fig. \ref{fig:Pix2Pix} illustrates the generator structure. CycleGAN also has the same U-Net generator, but the training scheme slightly differs from the Pix2Pix \cite{cyclegan}. The generator has eight down-sampling blocks and seven up-sampling blocks. In comparison, the discriminator consists of three down-sampling blocks followed by zero padding, convolution, batch normalization, leaky relu, and zero padding layers. The generator loss is the sum of the binary cross-entropy and the mean absolute loss multiplied by a constant \cite{pix2pix}. For the conversion of \ac{CT} to \ac{MRI} images, the \ac{GAN} model is trained on the dataset from \cite{dataset, dataset_2} with 75\% used on training and 25\% for testing.

\subsubsection{\textbf{Hardware Accelerator-Aware Reformulation}}

\begin{figure}[H]
    \centering
    \includegraphics[width=\linewidth]{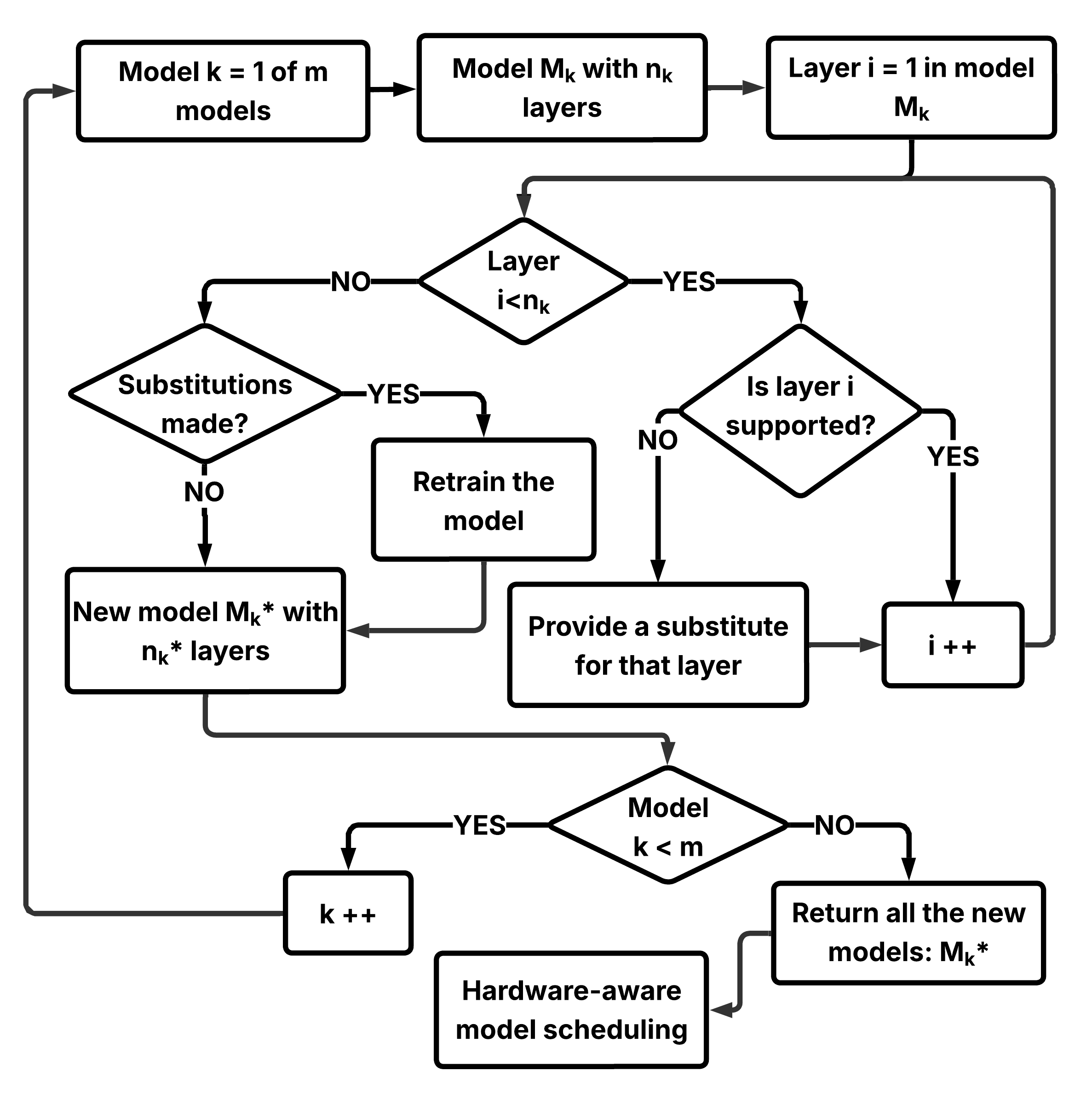}
    \caption{Methodology to produce a hardware-aware model}
    \label{fig:method}
\end{figure}

The methodology for reformulating a convolutional model to enable hardware-aware execution is illustrated in Fig. \ref{fig:method}. Each layer is first profiled to identify operations that are incompatible with the target hardware constraints. Following this, these identified layers are substituted with their hardware-compliant counterparts and retrained. 

The methodology for reformulating the convolutional model to enable hardware-aware execution is illustrated in Fig.~\ref{fig:method}. Each layer is first profiled to identify operations that are  The identified layers are then systematically replaced with hardware-compliant alternatives, after which the modified network is retrained to recover any performance degradation introduced by the substitutions.

The primary architectural incompatibility arises from transposed convolution (deconvolution) layers configured with non-zero padding. According to the \ac{DLA} constraints \cite{dla-limit}, these layers violate the requirements for \ac{DLA}- compliant layers, triggering GPU fallback during execution. The deconvolution layers swap the backward and forward passes of convolution. It is unclear what the exact reason for this incompatibility is. However, one possible hypothesis is that the padding for deconvolution uses a different computational pattern, requiring a different hardware implementation that has not been configured in the \ac{DLA}. When the padding operation is externalized and implemented as a separate layer, TensorRT interprets it as an independent operator within the computational graph. The library may treat it as a separate operation that can be executed as a layer on its own or may merge it with adjacent operations. Since it is treated independently, the reformulation may result in a longer latency due to additional memory transactions and kernel launches compared to when the padding is enabled on the deconvolution layer. 

\begin{figure*}[!t]
    \centering
    \includegraphics[width=\linewidth]{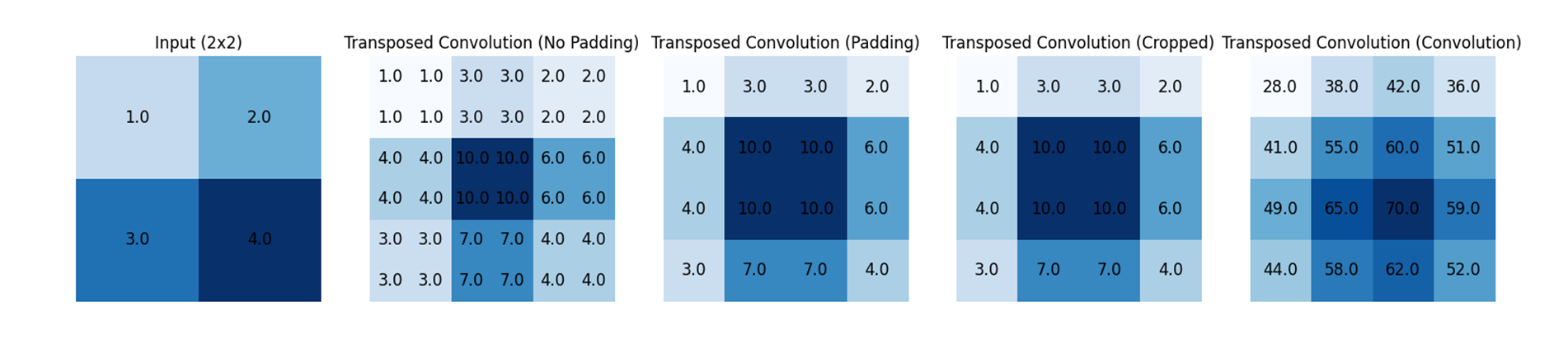}
    \caption{Input and outputs from the convolution transpose layer (deconvolution layer) with different replacements for the padding}
    \label{fig:input-output}
\end{figure*}

In the case of the Pix2Pix model, the padding trims the boundary of the output \cite{guide_to_convolution}. This requires some substitutions to ensure the model's integrity is not compromised while still allowing it to be fully implemented in the DLA. Equation \ref{eq:deconvolution} represents the relationship between the input and output size for the deconvolution layer. 
\begin{align} \label{eq:deconvolution}
    output\_size &= stride \cdot (input\_size - 1) + kernel\_size \notag \\
    & - 2 \cdot padding 
\end{align}
In the Pix2Pix model, the kernel size is 4, and the stride is 2 for all layers. Without padding, equation \ref{eq:deconvolution} can be simplified to equation \ref{eq:deconvolution_simplified_no_pad}. 
\begin{align} \label{eq:deconvolution_simplified_no_pad}
    output\_size &= 2 \cdot input\_size + 2 
\end{align}
With padding, the original equation is simplified to equation \ref{eq:deconvolution_simplified_pad}, where the output is double the size of the input. 
\begin{align} \label{eq:deconvolution_simplified_pad}
    output\_size &= 2 \cdot input\_size 
\end{align}
Fig. \ref{fig:input-output} reflects this difference between the deconvolution layer with and without padding. In this case, the padding operation effectively trims boundary pixels to preserve exact spatial doubling. Substitutions for the padding operation must reflect this change while being \ac{DLA} compatible. 

\textbf{1) Cropping Layer:} One possibility is using the cropping layer, which removes the specified number of rows/columns from the four borders. If the layer is set to remove a single row/column from the four borders (equation \ref{eq:cropping}), it is able to emulate the behaviour of the padding operation in the deconvolution layer. 

\begin{align} \label{eq:cropping}
    output\_size &= input\_size - 2 
\end{align}

\textbf{2) Convolution Layer:} An alternative solution is to use the convolution layer. Equation \ref{eq:convolution} represents the relationship between the input and output size for the convolution layer. 
\begin{align} \label{eq:convolution}
    output\_size &= \Bigl \lfloor \frac{1}{stride} ( input\_size - kernel\_size \notag \\
    &+ 2 \cdot padding ) \Bigr \rfloor  + 1  
\end{align}
If the kernel size is set to three, the stride to one, and there is no padding, the above equation can be simplified to equation \ref{eq:convolution_simplified}. 
\begin{align} \label{eq:convolution_simplified}
    output\_size &= input\_size - 2 
\end{align}
Similar to the cropping layer, the convolution layer can trim edges and produce results similar to those of the deconvolution layer with padding. Due to the nature of the convolutional layer, additional parameters are added, which can affect the values.  

\textbf{3) Other substitutions tested:} Apart from these two substitutions, other layers were also explored and are included below (in addition to removing padding from the deconvolution layers): 
\begin{itemize}
    \item Average pooling layer
    \item Maximum pooling layer
    \item Reduced kernel size of all deconvolution layers to two
    \item Reduced kernel size of all layers to two
    \item Removed padding from all convolution layers
\end{itemize}
These layers were proven to negatively impact the accuracy of the model and dramatically decreased performance compared to the original model. The cropping layer and convolution layer slightly outperformed the original model, therefore, becoming ideal substitutions for the padding operation that are \ac{DLA} compatible. Even though these substitutions came with an additional ten "unnamed" layers as a result of the dynamic inputs, the ONNX GraphSurgeon tool \cite{onnx-graph} eliminated these layers.

\subsection{Execution Pipelines and Scheduling Strategies}

This section explores possible execution pipelines and scheduling strategies. Performance metrics, transition layer analysis, throughput, utilization, and power consumption are discussed afterward in Section V.

\begin{figure}[H]
    \centering
    \includegraphics[width=\linewidth]{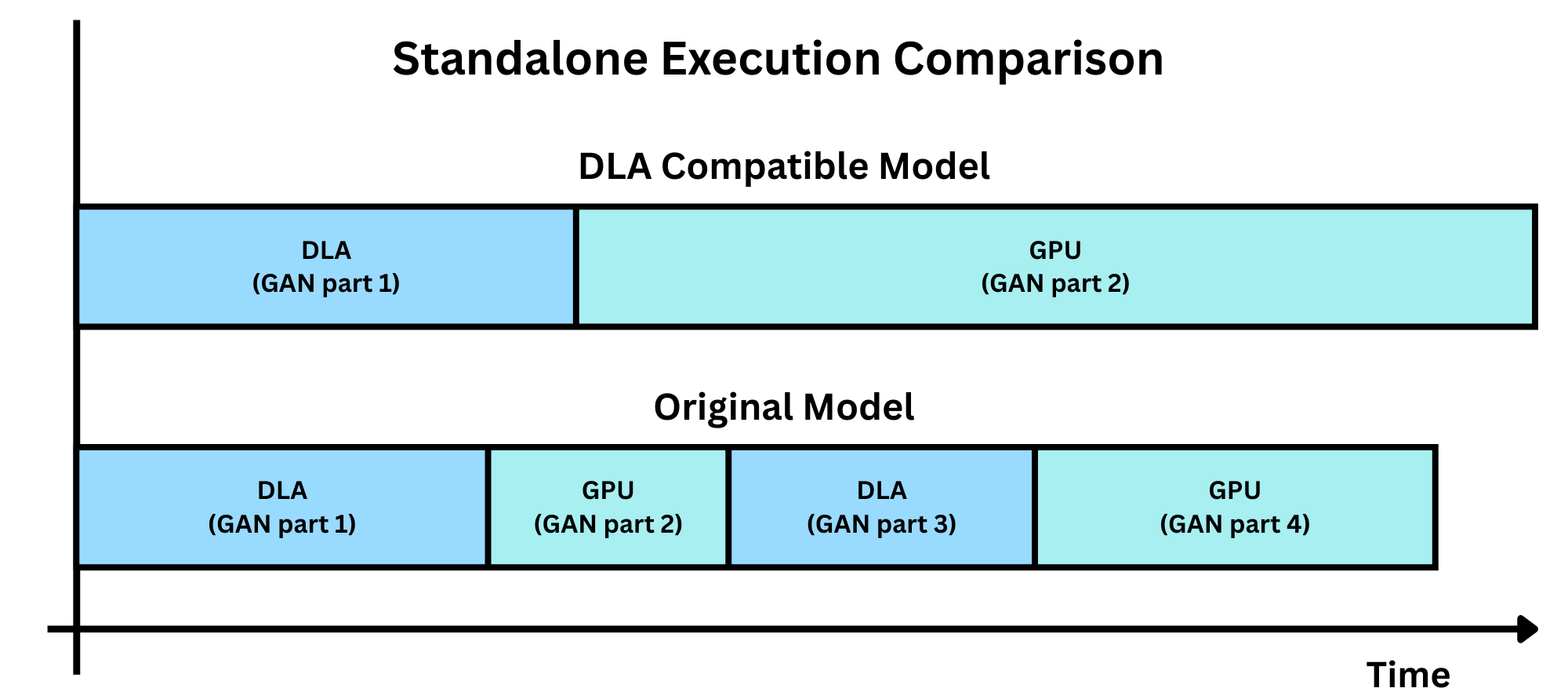}
    \caption{Execution of the original model and modified model in standalone scheme}
    \label{fig:standalone}
\end{figure}

\subsubsection{\textbf{Standalone Execution}}

The models were initially profiled in a standalone execution scheme. In this configuration, a single model is executed independently to isolate raw accelerator behavior without scheduling interference. The original model, despite being incompatible with the \ac{DLA}, executes with GPU fallback when unsupported layers are encountered. The modified models execute primarily on the \ac{DLA}, eliminating fallback through padding substitution. Fig. \ref{fig:standalone} represents this scheme.

\begin{figure}[H]
    \centering
    \includegraphics[width=\linewidth]{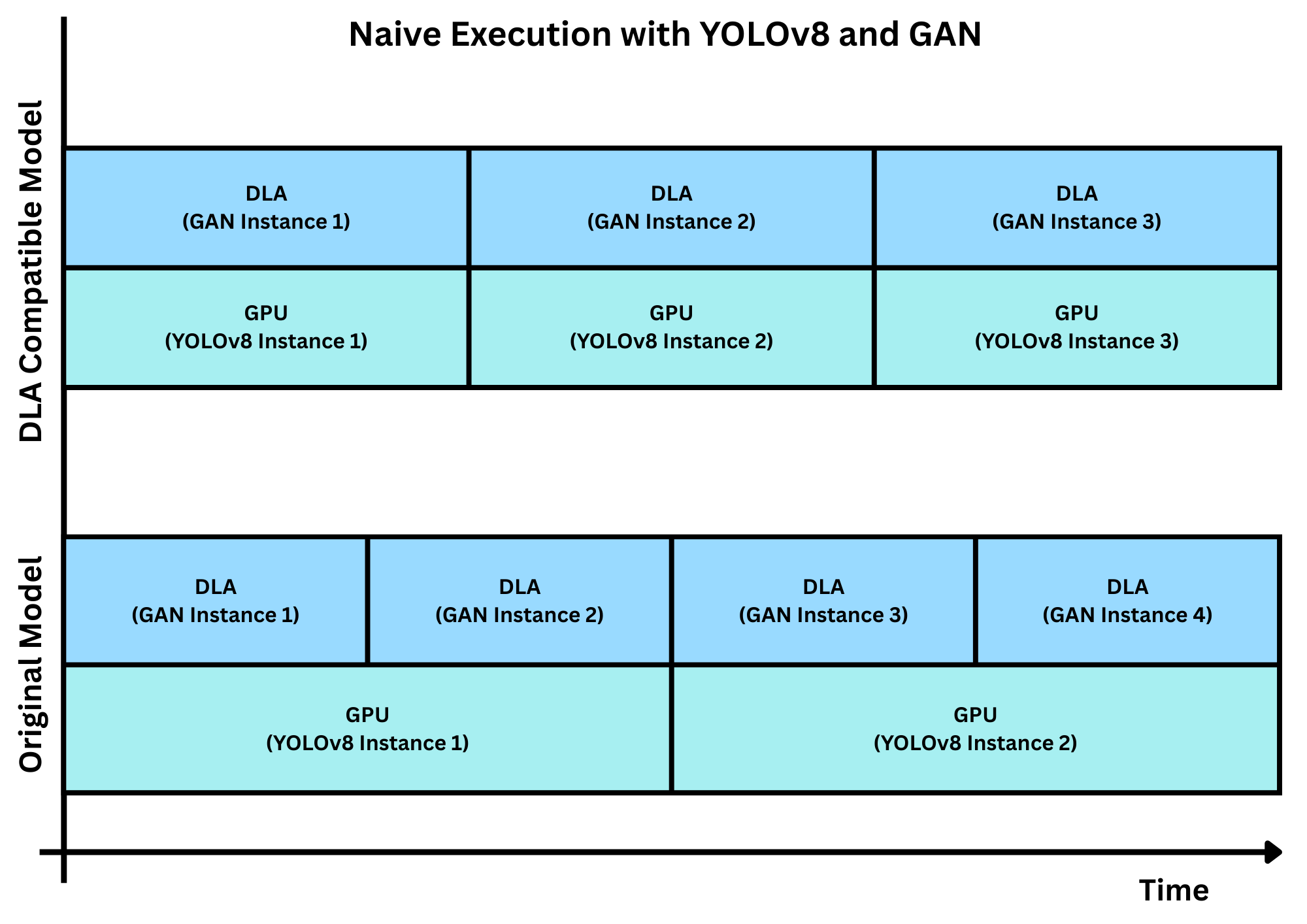}
    \caption{Difference between models in the timing diagram of the naive execution (server-level scheme)}
    \label{fig:server}
\end{figure}

\subsubsection{\textbf{Naive Concurrent Scheduling (Client-Server Scheme)}}

\begin{figure}[H]
    \centering
    \includegraphics[width=\linewidth]{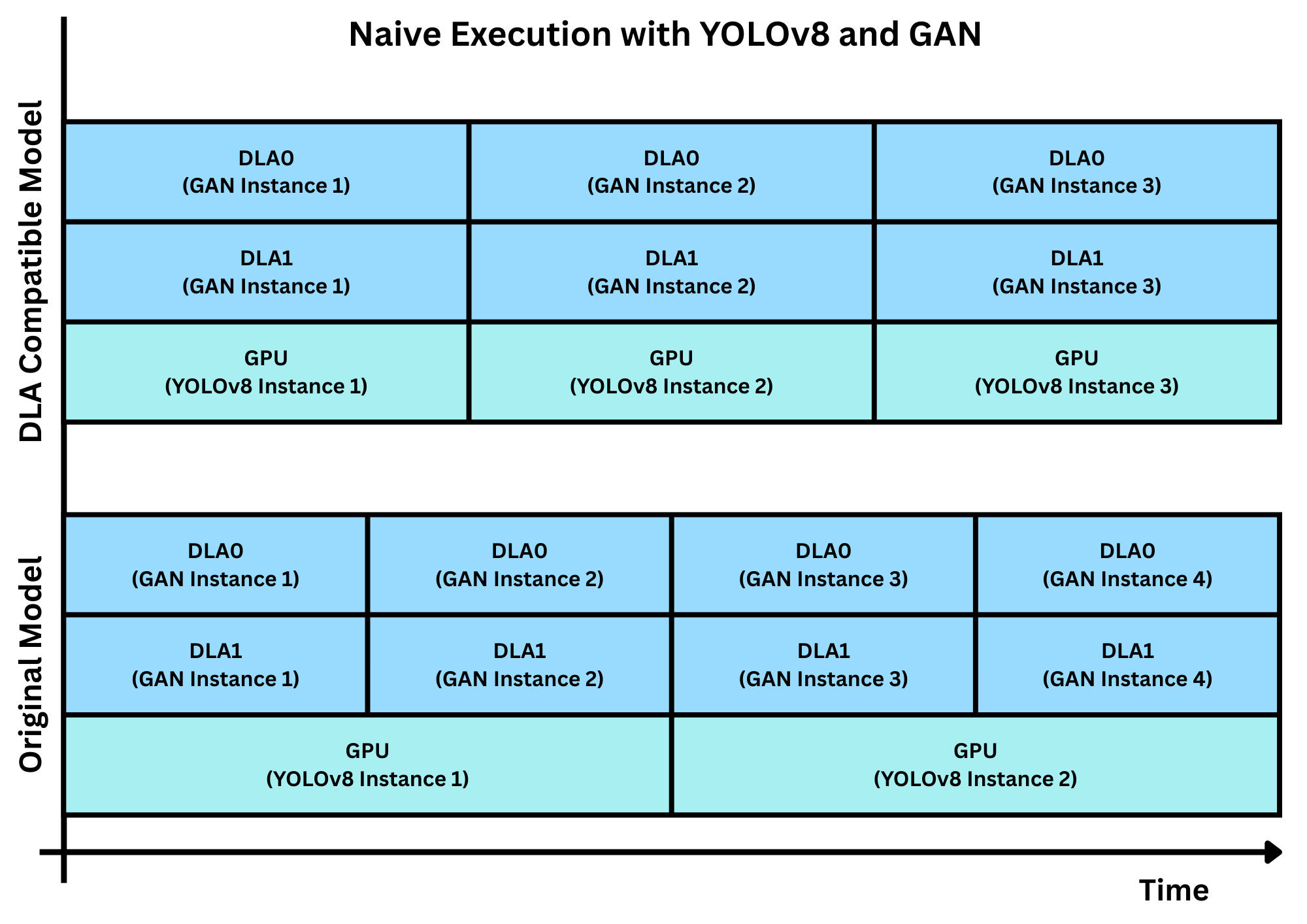}
    \caption{Difference between models in the timing diagram of the naive execution using both \ac{DLA} (server-level scheme)}
    \label{fig:2dla}
\end{figure}

The proposed naive methodology focuses on efficiently executing multiple deep learning models on heterogeneous edge devices for server-level medical imaging applications. The YOLO model is designed to detect strokes using machine learning. The \ac{GAN} model is trained to reconstruct \ac{MRI} images from \ac{CT} images. For a naive schedule, the \ac{GAN} model is executed on the \ac{DLA}, while the YOLOv8 model is executed on the GPU. In this approach, models are statically assigned to accelerators without layer partitioning. This type of scheduling may be ideal for server-level execution rather than real-time applications, as there may be considerable delay between frames for diagnosing and reconstruction. This methodology can also be extended to use both \acp{DLA}, with two instances of the GAN model running on DLA0 and DLA1, while the GPU concurrently runs the YOLOv8 model. Fig. \ref{fig:server} and \ref{fig:2dla} illustrate two naive execution strategies: one using a single DLA and the other using both DLAs.

\subsubsection{\textbf{HaX-CoNN Multi-Stream Scheduling (Real-Time Standalone Scheme)}}

\begin{figure}[H]
    \centering
    \includegraphics[width=\linewidth]{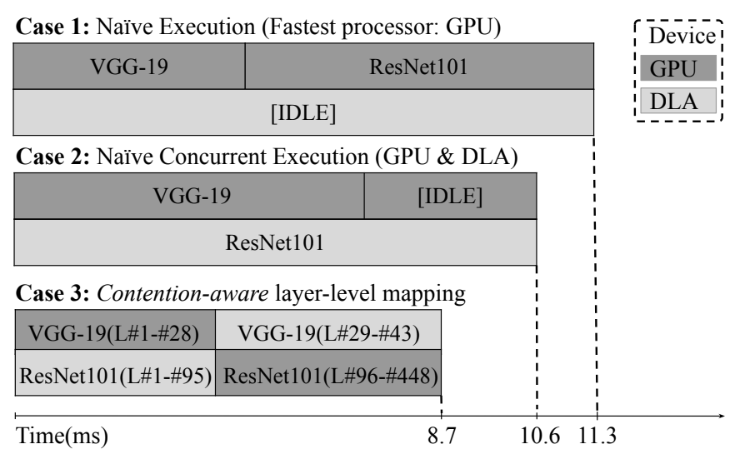}
    \caption{Timing diagram of \ac{HaX-CoNN} scheduling (case 3) compared to GPU-only (case 1) and GPU-DLA (case 2) \cite{HaX-CoNN}}
    \label{fig:HaX-CoNN_timing}
\end{figure}

To reduce delays and provide a more appropriate pipeline for real-time execution, we adopt a multi-streaming execution strategy inspired by the \ac{HaX-CoNN} scheduling technique \cite{HaX-CoNN}. In this approach, a \ac{SAT} solver ensures that two DNN models are executed concurrently without idle time by partitioning layers and swapping the model instances between the independent accelerators. In \ac{HaX-CoNN}, profiling data are used to pre-assign incompatible layers to the GPU, thereby preventing unsupported execution on the \ac{DLA} but simultaneously constraining the transition search space. In contrast, the proposed approach expands the set of \ac{DLA}-compatible layers, increasing the number of feasible partitioning configurations, which can lead to improved scheduling outcomes in terms of latency and accelerator utilization.

\begin{figure}[H]
    \centering
    \includegraphics[width=\linewidth]{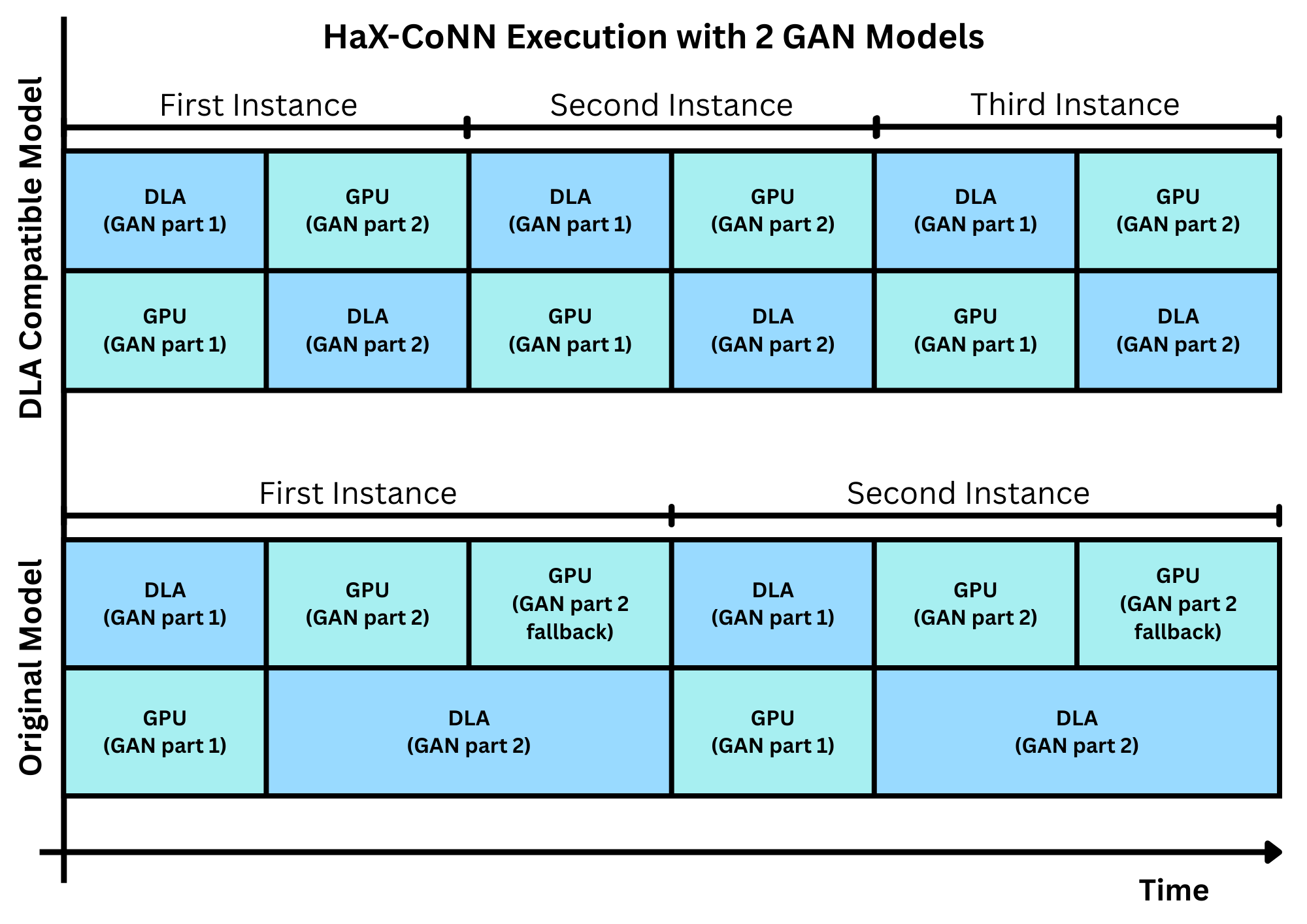}
    \caption{Difference between models in the timing diagram of the HaX-CoNN scheduling with two GAN instances}
    \label{fig:2GAN}
\end{figure}

\begin{figure}[H]
    \centering
    \includegraphics[width=0.9\linewidth]{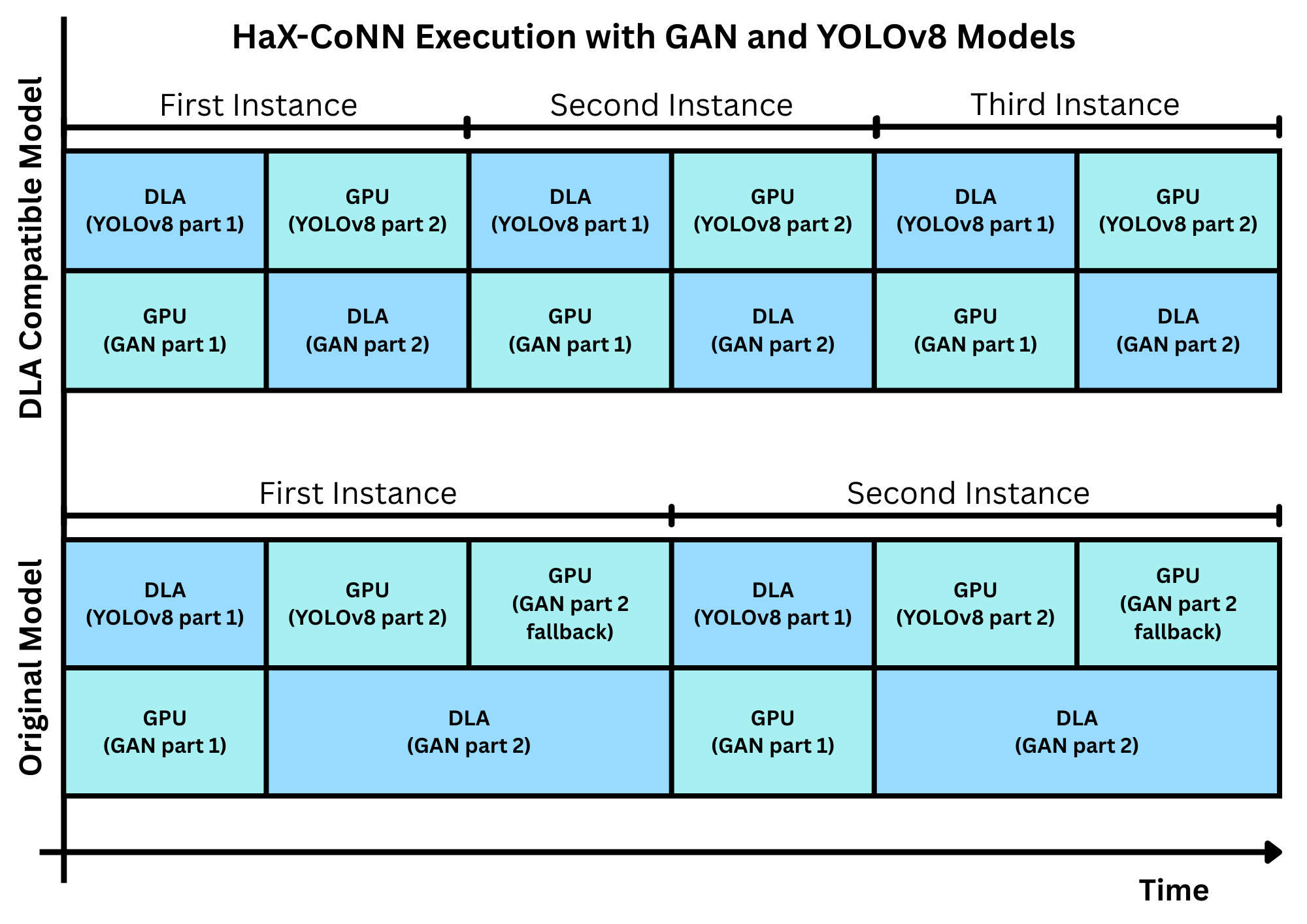}
    \caption{Difference between models in the timing diagram of the HaX-CoNN scheduling with GAN and YOLOv8 instances (standalone scheme)}
    \label{fig:GAN+YOLOv8}
\end{figure}

A straightforward scheduling heuristic can be derived by aligning the execution times of the GPU and the \ac{DLA} to balance workloads across devices. Using Fig. \ref{fig:HaX-CoNN_timing}, this means that the latency of layers 1 to 28 for VGG-19 (on the GPU) is close to the latency for the ResNet101 model for layers 1 to 95 (on the \ac{DLA}). Similarly, the latency of layers 29 to 43 for the VGG-19 (on the \ac{DLA}) is roughly the same as that of layers 96 to 448 for the ResNet101 model (on the GPU). This type of scheduling can be applied to different combinations of models, such as two instances of the GAN reconstruction model, or one GAN model for reconstruction and another for diagnostic detection. The hardware-aware model would ensure the pipeline operates at peak performance by eliminating GPU fallbacks. Two deployment scenarios are evaluated under HaX-CoNN scheduling: two concurrent GAN reconstruction models (Fig. \ref{fig:2GAN}), and a combined GAN reconstruction and YOLOv8 diagnostic detection pipeline representative of CT imaging systems (Fig. \ref{fig:GAN+YOLOv8}).

\section{Experimental Results and Discussion}

\subsection{Accuracy Comparison Between Original and Hardware-Aware Models}

After training, the models were evaluated using metrics such as \ac{PSNR}, \ac{SSIM}, and \ac{MSE}. Table \ref{accuracy} summarizes the results for the original Pix2Pix model and its hardware-aware variants. The cropping-based DLA model achieves an SSIM of 74.61 compared to 75.38 for the baseline, with nearly identical PSNR and closely aligned MSE. The convolution-based variant similarly maintains competitive performance, achieving an SSIM of 73.03 and a PSNR of 22.07 dB. Although this version introduces additional parameters to ensure DLA compatibility, the resulting reconstruction quality remains within a narrow margin of the baseline model. Overall, the results indicate that enforcing DLA compatibility through padding substitution preserves reconstruction fidelity within a narrow margin of the original architecture, while enabling improved hardware utilization.

A similar evaluation was conducted for CycleGAN. As shown in Table \ref{accuracy_cyclegan}, the cropping variant maintains reconstruction quality close to the original model, with SSIM remaining within an acceptable margin. While a moderate reduction in SSIM is observed, the structural similarity between reconstructed and target images remains within a reasonable margin for transformation-based GAN models. These results demonstrate that DLA-compatible reformulation preserves signal fidelity while enabling accelerator-level execution.

\begin{table}[!htbp]
\centering
\renewcommand{\arraystretch}{1.2}
\caption{Accuracy comparison between the original and modified Pix2Pix models}
\label{accuracy}
\begin{tabularx}{\columnwidth}{@{} m{1cm} *{3}{>{\centering\arraybackslash}X} @{}}
\toprule
\textbf{Value} & 
\makecell{\textbf{Original}\\\textbf{Pix2Pix}} & 
\makecell{\textbf{Pix2Pix with}\\\textbf{Cropping}} & 
\makecell{\textbf{Pix2Pix with}\\\textbf{Convolution}} \\ 
\midrule
Parameters & 54,425,859 & 54,425,859 & 64,637,268 \\ 
SSIM ↑     & 75.38 & 74.61 & 73.03 \\ 
PSNR ↑     & 23.12 & 22.99 & 22.07 \\ 
MSE ↓      & 38.28 & 38.04 & 39.83 \\
\bottomrule
\end{tabularx}
\end{table}

\begin{table}[!htbp]
\centering
\renewcommand{\arraystretch}{1.2}
\caption{Accuracy comparison between the original and hardware-aware CycleGAN models}
\label{accuracy_cyclegan}
\begin{tabularx}{\columnwidth}{@{} m{1.2cm} *{2}{>{\centering\arraybackslash}X} @{}}
\toprule
\textbf{Value} & 
\makecell{\textbf{Original}\\\textbf{CycleGAN}} & 
\makecell{\textbf{CycleGAN with}\\\textbf{Cropping}} \\ 
\midrule
SSIM ↑ & 48.24 & 42.96 \\ 
PSNR ↑ & 18.83 & 18.96 \\ 
MSE ↓  & 50.09 & 49.36 \\
\bottomrule
\end{tabularx}
\end{table}

\subsection{Data Collection}

Timing analytics were collected using the trtexec utility from TensorRT for standalone profiling \cite{tensorrt} and the DeepStream SDK \cite{deepstream} for concurrent profiling. The timing diagrams were taken from the NVIDIA Nsight Systems \cite{nsight}. Power statistics were collected using the tegrastats utility, which records the voltage, current, power, utilization, and frequency of the CPU, GPU, and other accelerators on the device \cite{power}. During preliminary measurements, it was observed that the execution order of the models influenced the reported power consumption, even when substantial idle intervals were introduced between runs. Increasing the number of trials converges the power values across the different models, ultimately revealing negligible differences between the hardware-aware and original implementations. This outcome may result from the continuous optimization of hardware resources to produce ideal execution. The results shown in this paper later on were taken after the machine was rebooted for each model. It is also important to note that prior studies have reported that tegrastats may underestimate absolute power consumption \cite{power-better}.

\subsection{Performance in Standalone Execution}

\begin{figure}[H]
    \centering
    \includegraphics[width=\linewidth]{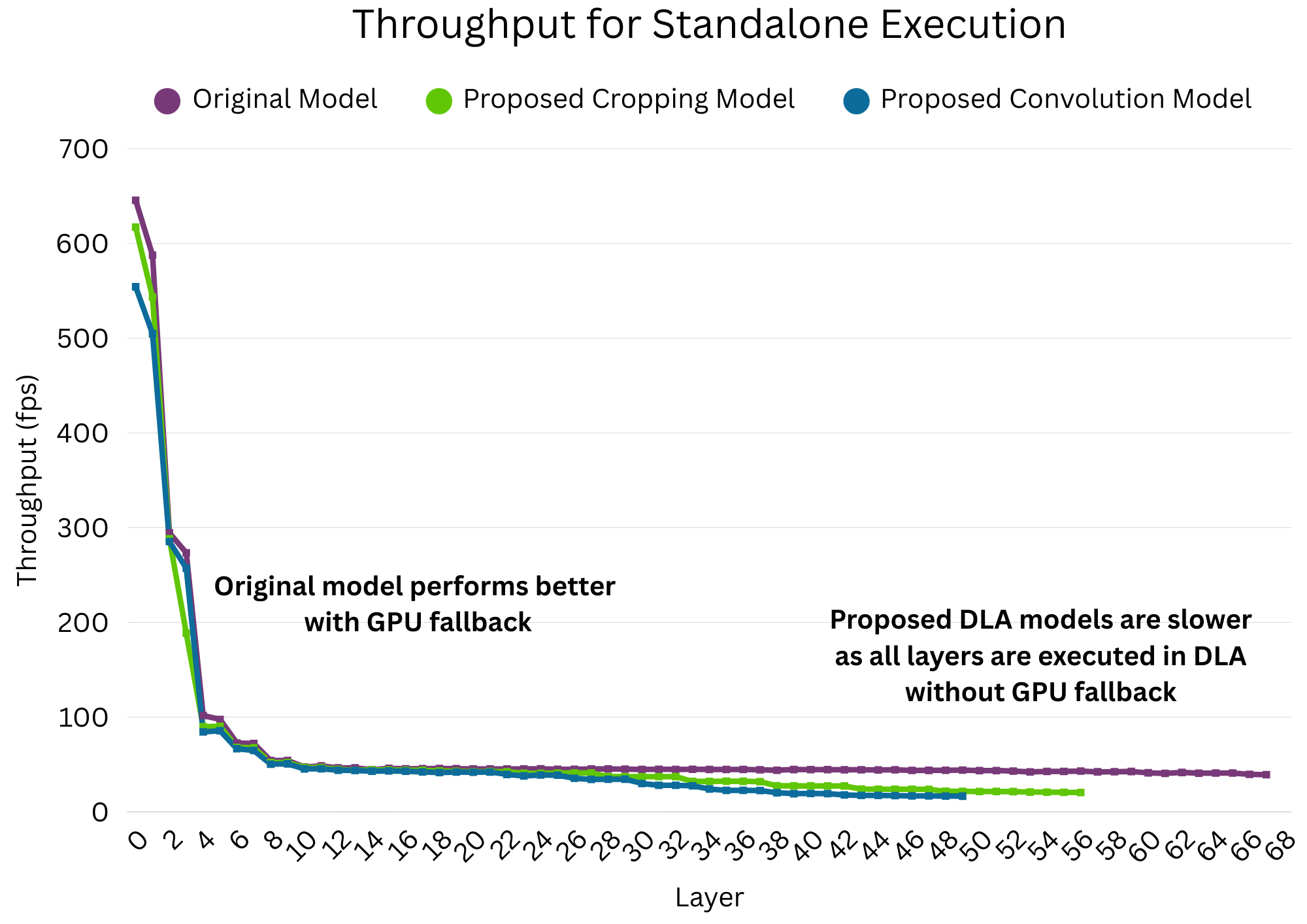}
    \caption{Throughput for the standalone execution}
    \label{fig:throughput_standalone}
\end{figure} 

\begin{figure}[H]
    \centering
    \includegraphics[width=\linewidth]{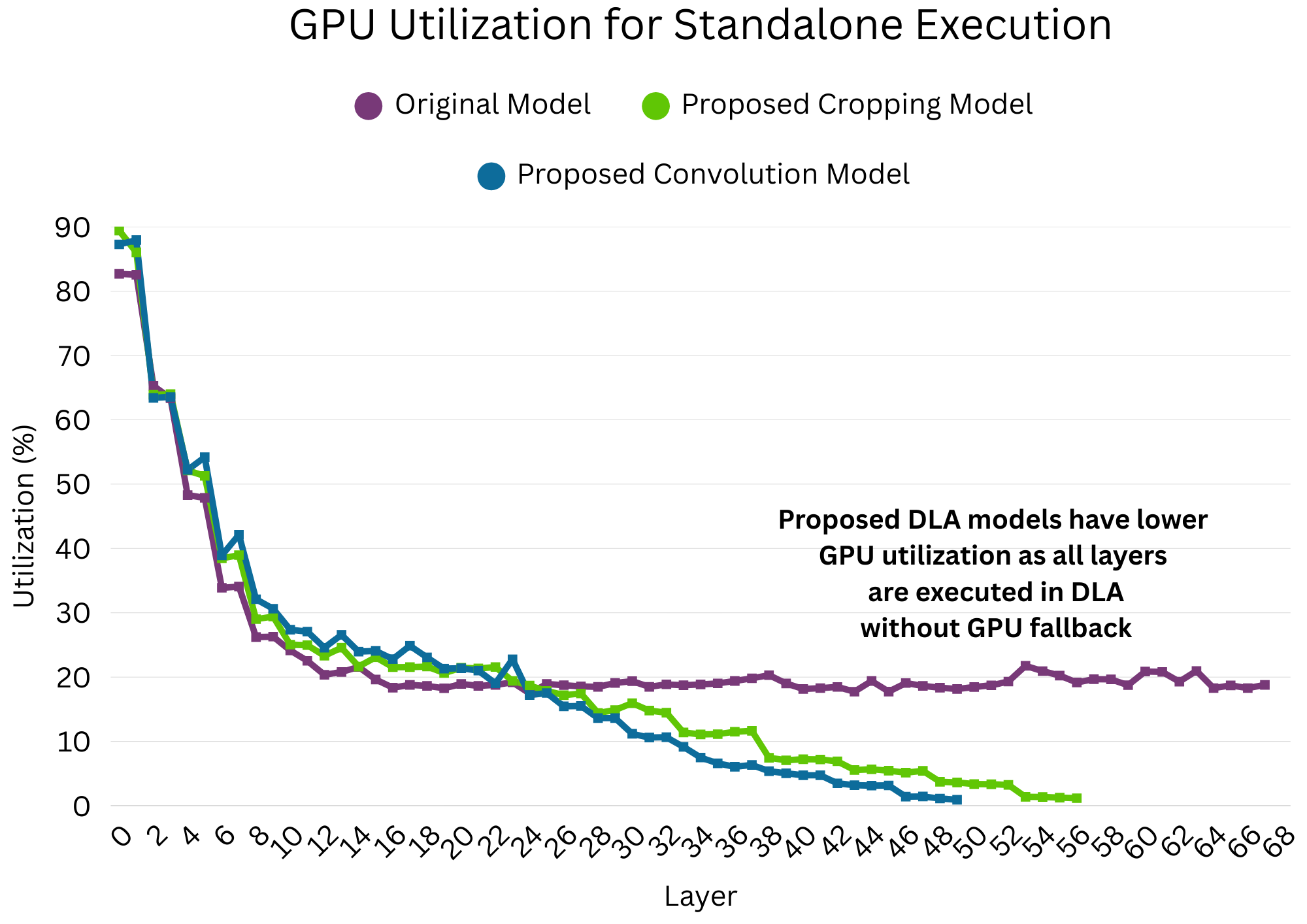}
    \caption{GPU Utilization for the standalone execution}
    \label{fig:utilization_standalone}
\end{figure}

The models were initially profiled in a standalone execution (as shown in Fig. \ref{fig:standalone}). The original model, despite being incompatible with the \ac{DLA}, achieves higher throughput than the two modified versions, as shown in Fig. \ref{fig:throughput_standalone}. This performance difference can be attributed to increased operations compared to the original model. Although the proposed cropping model is similar to the original model, because the padding (cropping layer) operation is separated from the convolution, it is treated as an operation on its own, increasing latency and, therefore, reducing throughput. In the case of the proposed convolution model, the number of convolution operations have increased, requiring more computation time. Additionally, the increase in latency can prove that the latency incurred from multiple transitions between the \ac{DLA} and GPU due to incompatible layers is less than their execution time in the \ac{DLA} when modified. The effect of the substitution is more prominent when examining the utilization data in Fig. \ref{fig:utilization_standalone}. As more layers are added to the \ac{DLA}, the modified models can minimize GPU utilization to zero, whereas the original model remains around 20\%. Thus, replacing the deconvolution padding successfully eradicated GPU fallback.

\subsection{Performance in Naive Concurrent Scheduling}

As discussed previously, for a naive schedule, the \ac{GAN} model is executed in the \ac{DLA} with the YOLOv8 model executed in the GPU, as presented in Fig. \ref{fig:server}. This methodology is also extended to utilize both \acp{DLA}, as shown in Fig. \ref{fig:2dla}. Both \acp{DLA} execute two instances of the GAN model while the GPU executes an instance of the YOLOv8 model. The \ac{DLA}-incompatible layers in the original model interrupt GPU execution of the YOLOv8 model, hindering performance and reducing throughput. The hardware-aware models improve throughput by 23\%-25\%, demonstrating the detrimental effect of GPU fallback on concurrent execution. On the other hand, the throughput of the \ac{DLA} is consistent with that of the standalone execution, where the original model can outperform the modified models (Table \ref{results_for_naive}). With the reduced fallback, the GPU is able to make effective use of the \acp{SM}, resulting in a greater GPU utilization for the proposed models (91\%) compared to the original model (77\%). The increased utilization does, however, cause an increase in power consumption (Table \ref{power_results_naive}). Similarly, using both \ac{DLA} shows a similar trend in throughput, utilization, and power consumption, as seen in Tables \ref{results_for_naive_2} and \ref{power_results_naive_2}.

\begin{table}[!htbp]
\centering
\renewcommand{\arraystretch}{1.2}
\caption{Throughput of YOLOv8 and Pix2Pix models under naive execution}
\label{results_for_naive}
\begin{tabular}{@{} l c c @{}}
\toprule
\textbf{Model} & \textbf{GPU (FPS)} & \textbf{DLA (FPS)}\\ 
\midrule
Original Pix2Pix & 326.88 & 61.39 \\ 
Proposed Pix2Pix (Cropping Layer) & 402.45 & 29.39 \\ 
Proposed Pix2Pix (Convolution Layer) & 407.28 & 23.71 \\ 
\bottomrule
\end{tabular}
\end{table}

\begin{table}[!htbp]
\centering
\renewcommand{\arraystretch}{1.15}
\caption{Hardware utilization and power consumption of YOLOv8 and Pix2Pix models under naive execution}
\label{power_results_naive}
\begin{tabular}{@{} l c c @{}}
\toprule
\textbf{Model} & 
\textbf{GPU Util (\%)} & 
\textbf{Power (mW)} \\ 
\midrule
Original Pix2Pix & 76.50 & 26377.73 \\ 
Proposed Pix2Pix (Cropping Layer) & 91.93 & 33317.82 \\ 
Proposed Pix2Pix (Convolution Layer) & 91.38 & 34248.66 \\ 
\bottomrule
\end{tabular}
\end{table}

\begin{table}[!htbp]
\centering
\renewcommand{\arraystretch}{1.2}
\caption{Throughput of YOLOv8 and Pix2Pix models under naive execution using both \ac{DLA}}
\label{results_for_naive_2}
\begin{tabular}{@{} l c c c @{}}
\toprule
\textbf{Model} & \textbf{GPU (FPS)} & \textbf{DLA0 (FPS)} & \textbf{DLA1 (FPS)}\\ 
\midrule
Original Pix2Pix & 262.41 & 59.51 & 58.98\\ 
Proposed Pix2Pix \\ (Cropping Layer) & 361.74 & 29.04 & 28.53\\ 
Proposed Pix2Pix \\ (Convolution Layer) & 362.71 & 23.38 & 23.01\\ 
\bottomrule
\end{tabular}
\end{table}

\begin{table}[!htbp]
\centering
\renewcommand{\arraystretch}{1.15}
\caption{Hardware utilization and power consumption of YOLOv8 and Pix2Pix models under naive execution using both \ac{DLA}}
\label{power_results_naive_2}
\begin{tabular}{@{} l c c @{}}
\toprule
\textbf{Model} & 
\textbf{GPU Util (\%)} & 
\textbf{Power (mW)} \\ 
\midrule
Original Pix2Pix & 78.29 & 23445.20 \\ 
Proposed Pix2Pix (Cropping Layer) & 89.04 & 27472.53 \\ 
Proposed Pix2Pix (Convolution Layer) & 88.71 & 23672.16 \\ 
\bottomrule
\end{tabular}
\end{table}

\subsection{Performance in HaX-CoNN Scheduling}

\subsubsection{\textbf{Two \ac{GAN} MRI Reconstruction Models}} 
Using the TensorRT profiling data \cite{tensorrt} for different transition layers, a tentative schedule was designed by correlating the measured \ac{DLA} and GPU latencies for the \ac{GAN} models. The resulting hardware-aware allocation strategy is summarized in Table \ref{schedule}. The table outlines the different partitioning layers for each model. For example, in the original model, the first instance assigns the first four layers to the DLA and the remaining layers to the GPU, whereas the second instance assigns the first 10 layers to the GPU and the remaining layers to the DLA. For the DLA-compatible model, the partition occurs after the first five layers on the DLA, with the first 51 or 42 layers on the GPU, depending on the type of substitution (cropping or convolution, respectively). Once the schedule was identified, the DeepStream library \cite{deepstream} was used to implement the pipeline. Fig. \ref{fig:2GAN} shows the timing diagram and how DLA-compatible models can improve performance by eliminating fallback. The throughput results are presented in Table \ref{results_for_HaX-CoNN}, while the hardware utilization and power statistics are summarized in Table \ref{power_results}. 

\begin{table}[!htbp]
\centering
\renewcommand{\arraystretch}{1.2}
\caption{Partitioning point for each of the Pix2Pix model when executed in HaX-CoNN manner}
\label{schedule}
\begin{tabular}{@{} l c c @{}}
\toprule
\textbf{Model} & \textbf{DLA to GPU} & \textbf{GPU to DLA}\\ 
\midrule
Original Pix2Pix & 4 & 10 \\ 
Proposed Pix2Pix (Cropping Layer) & 5 & 51 \\ 
Proposed Pix2Pix (Convolution Layer) & 5 & 42 \\ 
\bottomrule
\end{tabular}
\end{table}

\begin{table}[!htbp]
\centering
\renewcommand{\arraystretch}{1.2}
\caption{Throughput of Pix2Pix models under HaX-CoNN execution}
\label{results_for_HaX-CoNN}
\begin{tabular}{@{} l c c @{}}
\toprule
\textbf{Model} & \textbf{GPU (FPS)} & \textbf{DLA (FPS)}\\ 
\midrule
Original Pix2Pix & 160.29 & 76.75 \\ 
Proposed Pix2Pix (Cropping Layer) & 158.36 & 141.20 \\ 
Proposed Pix2Pix (Convolution Layer) & 129.03 & 129.02 \\ 
\bottomrule
\end{tabular}
\end{table}

\begin{table}[!htbp]
\centering
\renewcommand{\arraystretch}{1.15}
\caption{Hardware utilization and power consumption of Pix2Pix models under HaX-CoNN execution}
\label{power_results}
\begin{tabular}{@{} l c c @{}}
\toprule
\textbf{Model} & 
\textbf{GPU Util (\%)} & 
\textbf{Power (mW)} \\ 
\midrule
Original Pix2Pix & 59.28 & 24999.07 \\ 
Proposed Pix2Pix (Cropping Layer) & 63.41 & 23445.42 \\ 
Proposed Pix2Pix (Convolution Layer) & 72.33 & 28648.56 \\ 
\bottomrule
\end{tabular}
\end{table}

The original model achieves the highest peak GPU throughput (160.29 FPS); however, the DLA throughput drops significantly (76.75 FPS), indicating an imbalance caused by GPU fallback and fragmented execution. This imbalance is reflected in the Nsight timing diagrams (Fig. \ref{fig:Nsight}), where shorter and irregular DLA execution blocks introduce idle periods between accelerators. In contrast, the hardware-aware cropping configuration demonstrates a more balanced execution profile, achieving 158.36 FPS and 141.20 FPS across the two model instances, while reducing GPU utilization and significantly lowering total system power from 23.45 W to 25.00 W. The increased GPU utilization indicates improved workload distribution across the heterogeneous processors. Although the convolution configuration maintains symmetric throughput (129 FPS for both instances), it exhibits the highest GPU utilization (72.33\%) and total power consumption (28.65 W) due to the increased number of convolution operations. 

Overall, the cropping hardware-aware reformulation provides the best trade-off between throughput balance and energy efficiency. By eliminating GPU fallback and enabling sustained parallel execution, the system achieves near-symmetric FPS across instances while reducing total power consumption by approximately 6\% compared to the original model. This confirms that hardware-aware model reformulation not only improves scheduling stability but also enhances overall energy-performance efficiency on heterogeneous edge platforms.

\subsubsection{\textbf{\ac{GAN} MRI Reconstruction and YOLOv8 Models}}
The YOLOv8 model \cite{ultralytics}, developed by Ultralytics, is a one-stage detector that modernizes the YOLO family with C2f blocks in the backbone, a PAN/FPN-style neck for multi-scale fusion, and an anchor-free split head that predicts centers and distances instead of anchor offsets, improving both speed and localization on small targets. It has been used in medical imaging to detect breast cancer \cite{breast-tumor, breast-tumor-2} as well as polyps in colonoscopy \cite{polyps}. The model was trained on the dataset \cite{yolo-dataset} to identify whether a patient is suffering from a stroke.

\begin{figure*}[!t]
    \centering
    \includegraphics[width=0.85\linewidth]{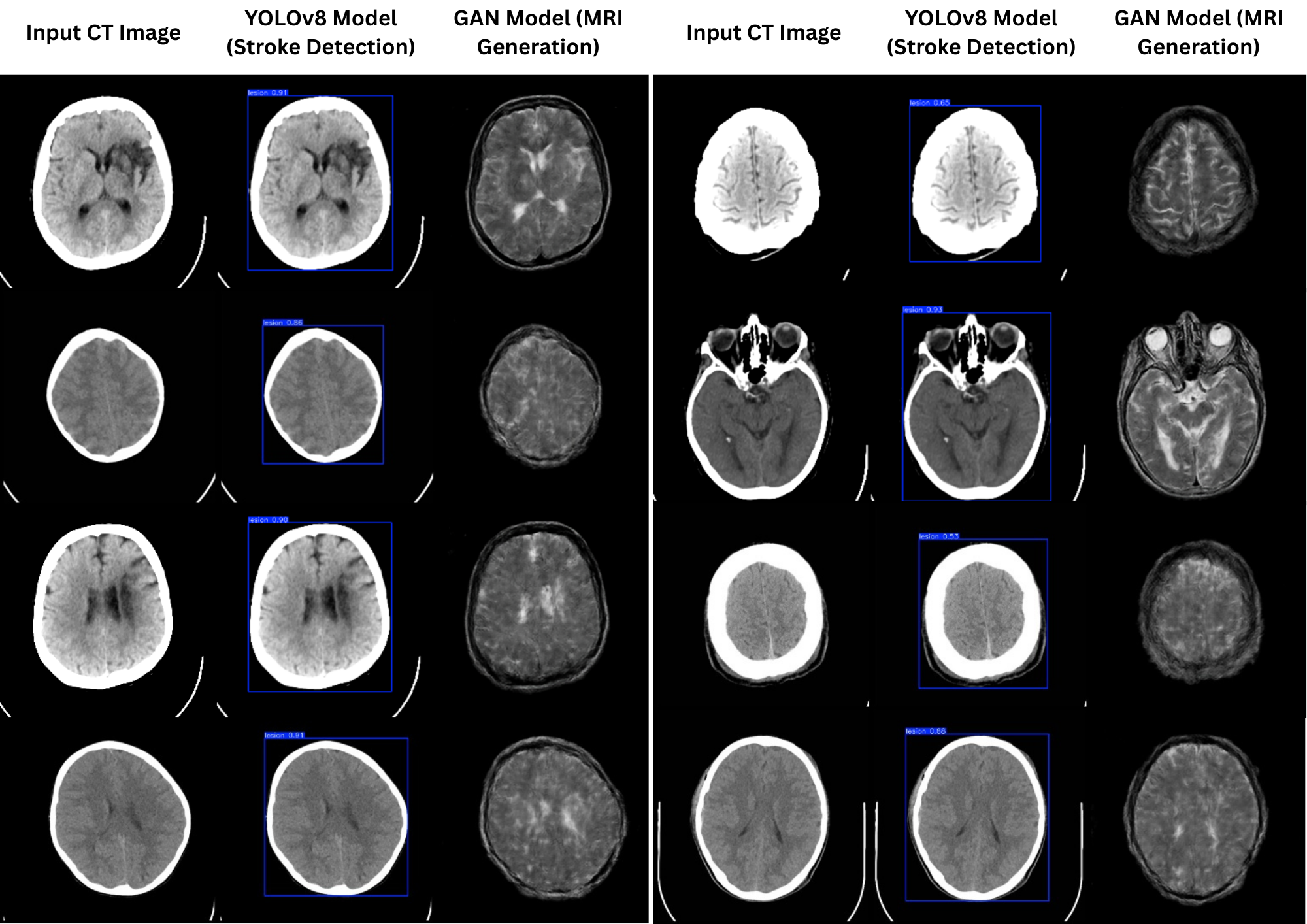}
    \caption{Input and output images from the two models}
    \label{fig:output_images}
\end{figure*}

Fig. \ref{fig:output_images} shows sample output images from the YOLOv8 model and the MRI generation with the Pix2Pix model. While the previous section focused on \ac{GAN}-only MRI reconstruction, this experiment demonstrates how the scheduling strategy extends to multi-model pipelines, specifically combining a GAN reconstruction model with a YOLOv8 diagnostic detection model. This scenario is representative of functional CT imaging setups, where both reconstruction and diagnostic classification occur concurrently on the same embedded platform. The same methodology was applied to identify optimal partition points for both models. Table \ref{schedule_recons} summarizes the scheduling decisions for the GAN-based reconstruction tasks. Fig. \ref{fig:GAN+YOLOv8} represents the timing diagram and how the elimination of fallback improves the performance of the pipeline. This system may be used as an addition to a CT scanner as presented in Fig. \ref{fig:schemes}A.

\begin{table}[!htbp]
\centering
\renewcommand{\arraystretch}{1.2}
\caption{Partitioning point for each of the Pix2Pix model and YOLOv8 model when executed in HaX-CoNN manner}
\label{schedule_recons}
\begin{tabular}{@{} l c c @{}}
\toprule
\textbf{Model} & \textbf{DLA to GPU} & \textbf{GPU to DLA}\\ 
\midrule
Original Pix2Pix & 7 & 10 \\ 
With Cropping Layer & 4 & 49 \\ 
With Convolution Layer & 6 & 43 \\ 
\bottomrule
\end{tabular}
\end{table}

\begin{table}[!htbp]
\centering
\renewcommand{\arraystretch}{1.2}
\caption{Throughput of each device for the Pix2Pix model with the YOLOv8 model when executed in HaX-CoNN manner}
\label{results_for_HaX-CoNN_recons}
\begin{tabular}{@{} l c c @{}}
\toprule
\textbf{Model} & \textbf{GPU (FPS)} & \textbf{DLA (FPS)}\\ 
\midrule
Original Pix2Pix & 127.31 & 86.73 \\ 
With Cropping Layer & 174.10 & 174.00 \\ 
With Convolution Layer & 166.89 & 166.85 \\ 
\bottomrule
\end{tabular}
\end{table}

\begin{table}[!htbp]
\centering
\renewcommand{\arraystretch}{1.15}
\caption{Hardware utilization and power consumption of the Pix2Pix model with the YOLOv8 model under HaX-CoNN execution}
\label{power_results_YOLOv8}
\begin{tabular}{@{} l c c @{}}
\toprule
\textbf{Model} & 
\textbf{GPU Util (\%)} & 
\textbf{Power (mW)} \\ 
\midrule
Original Pix2Pix & 45.81 & 21143.13 \\ 
Proposed Pix2Pix (Cropping Layer) & 74.62 & 26194.39 \\ 
Proposed Pix2Pix (Convolution Layer) & 70.50 & 28249.98 \\ 
\bottomrule
\end{tabular}
\end{table}

The performance outcomes for this combined pipeline are presented in Table \ref{results_for_HaX-CoNN_recons}. For the original Pix2Pix model, the GPU achieves 127.31 FPS while the \ac{DLA} reaches 86.73 FPS, resulting in a throughput imbalance of approximately 40.6 FPS. This disparity confirms that fallback-induced execution on the GPU prevents balanced concurrent scheduling. In the worst-case scenario, the original model achieves an overall throughput of 86.73 FPS. In contrast, the proposed hardware-aware models achieve a negligible difference between the GPU and DLA throughput, with the cropping model achieving 174 FPS, and the convolution-based reformulation reaching 166 FPS. These results demonstrate near-perfect throughput alignment between accelerators once \ac{DLA} incompatibilities are eliminated. Compared to the original model, the cropping-based variant improves GPU throughput by a factor of 2 (from 86.73 FPS to 174.10 FPS), increasing the effective \ac{DLA} utilization relative to its standalone constrained execution. The convolution-based model achieves a similar improvement in GPU throughput over the baseline. This balance is critical in concurrent multi-model pipelines, as it prevents bottlenecks and maximizes parallel hardware utilization.

Table \ref{power_results_YOLOv8} presents the corresponding GPU utilization and total power consumption. The original model operates at 45.81\% GPU utilization and draws 21 W. After hardware-aware reformulation, GPU utilization increases to 74.62\% for the cropping-based model and 70.50\% for the convolution-based model. This increase in utilization reflects improved parallel execution and reduced idle time caused by fallback transitions. The cropping-based model consumes 26 W, while the convolution-based model consumes 28 W. Although total power increases relative to the baseline, the substantial throughput gains (over 30\%) result in improved performance-per-watt efficiency for the hardware-aware configurations.

\begin{figure*}[!t]
    \centering
    \includegraphics[width=0.9\linewidth]{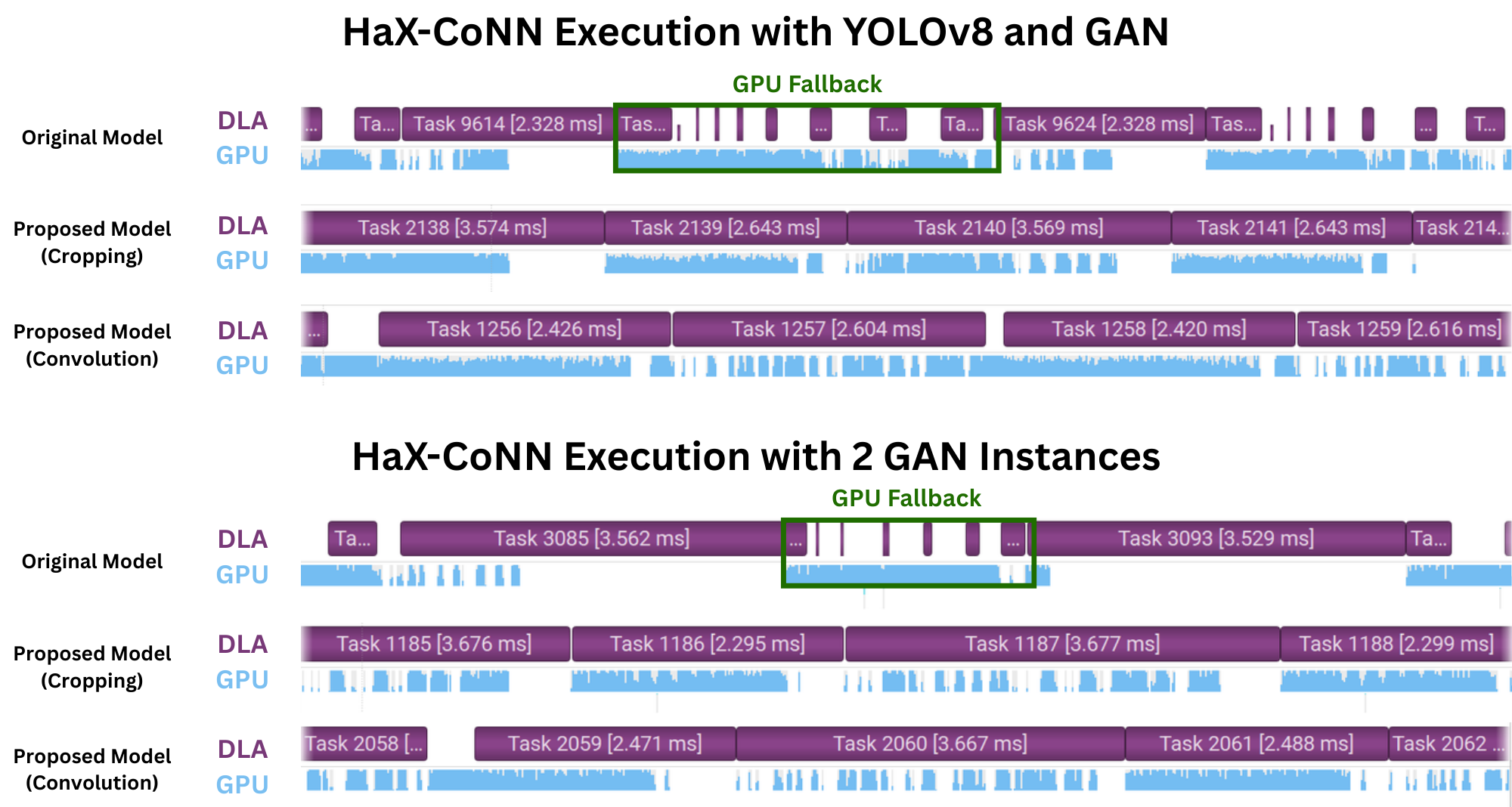}
    \caption{Difference between models in the Nsight Systems timing diagram of the real-time execution}
    \label{fig:Nsight}
\end{figure*}

This behavior is further confirmed in Fig.~\ref{fig:Nsight}, where the Nsight timing diagrams show synchronized and continuous execution blocks between the GPU and \ac{DLA} for the modified models. In contrast, the original model exhibits fragmented execution segments and visible idle gaps on the \ac{DLA}, corresponding to synchronization delays triggered by fallbacks. The visual alignment of execution intervals in the hardware-aware variants directly reflects the throughput symmetry observed in Table \ref{results_for_HaX-CoNN_recons}.

\section{Discussion and Future Prospects}

The experimental results confirm that the primary performance gains stem from the structural modification of \ac{DLA}-incompatible layers rather than from scheduling alone. In particular, replacing the padding operation within convolution transpose layers with functionally equivalent, \ac{DLA}-compatible operations (cropping or convolution-based trimming) eliminates GPU fallback and restores full accelerator residency. This seemingly localized architectural adjustment has system-level consequences: it removes fragmented execution, reduces subgraph generation, and enables predictable partitioning across heterogeneous processors.

The absence of fallback not only improves throughput but also stabilizes concurrent execution. In the original model, transitions between \ac{DLA} and GPU introduce irregular timing behavior and imbalance between accelerators. By enforcing full \ac{DLA} compatibility at the layer level, the modified models achieve balanced throughput across devices and enable deterministic scheduling under both naive and HaX-CoNN schemes. Furthermore, reducing the number of generated subgraphs lowers the risk of exceeding engine plan limits, thereby improving deployment robustness in multi-model scenarios.

Importantly, these improvements were achieved without degrading reconstruction quality. This demonstrates that hardware-aware layer refinement can preserve, and in some cases, enhance model fidelity while satisfying strict accelerator constraints.

Future investigations should extend this methodology beyond deconvolution padding. Other operator-level incompatibilities, including tensor reshaping constraints, dilation restrictions, and precision limitations, should be systematically analyzed and reformulated into accelerator-compliant equivalents. Developing a formalized framework for hardware-aware layer transformation could generalize this approach across convolutional and transformer-based architectures.

\subsection{AI Model Architectures}

As neural network architectures continue to increase in structural complexity, accelerator incompatibility will become a more critical bottleneck than raw computational throughput. The results suggest that compatibility must be treated as a first-class architectural constraint during model design rather than as a post-training optimization step.

Future work should therefore explore accelerator-aware neural architecture design, in which layers are constructed to satisfy hardware constraints. For example, deconvolution layers could be systematically replaced with upsampling–convolution combinations that respect padding and stride restrictions. Similarly, attention mechanisms and transformer blocks may require reformulation to avoid unsupported dynamic tensor operations or high-rank reshaping.

Rather than adapting trained networks retrospectively, co-design methodologies should integrate hardware constraints directly into the search space of neural architecture search or pruning pipelines. This would ensure that emerging models remain fully executable on fixed-function accelerators without fallback, enabling consistent multi-model concurrency. Another improvement to consider is the development of scheduling strategies for more than two models simultaneously. Currently, based on the review \cite{my_review}, \ac{LP} \cite{LP} is the only method; however, it has not been fully evaluated with respect to throughput, power consumption, and other metrics. \ac{HaX-CoNN} has only been tested for a single \ac{DLA}. 

\subsection{Hardware Improvements}

While model-level modifications resolve current compatibility limitations, architectural enhancements to the \ac{DLA} could further reduce the need for such transformations. Increasing on-chip buffer capacity would allow larger feature maps and intermediate tensors to remain fully resident within the accelerator, minimizing memory traffic and context switching. Additionally, implementing a double-buffering mechanism could reduce idle time during data movement, thereby improving pipeline utilization. Relaxing certain operator constraints, such as limited padding configurations or kernel parameter bounds, would further expand the class of natively supported models. However, until such hardware extensions become available, systematic layer reformulation remains a practical and scalable solution for achieving end-to-end accelerator residency.

\section{Conclusion}

This work demonstrated that targeted layer-level modifications can transform a partially accelerator-compatible model into a fully \ac{DLA}-resident architecture without sacrificing accuracy. By reformulating the padding operation within convolution transpose layers, GPU fallback was eliminated, enabling deterministic partitioning and balanced concurrent execution across heterogeneous processors. When combined with HaX-CoNN scheduling for real-time stroke detection and MRI reconstruction, the hardware-aware models achieved approximately 170 frames per second, doubling performance over the original model while maintaining balanced GPU and \ac{DLA} throughput. The modified models preserve the predictive performance of the original architecture, with no statistically significant degradation in reconstruction quality. These findings underscore a key principle: accelerator compatibility must be engineered at the operator level to unlock the full benefits of heterogeneous scheduling. Rather than relying solely on runtime schedulers to compensate for incompatibilities, structural refinement of neural network layers provides a more stable and scalable pathway toward high-throughput, low-power edge AI systems. Moreover, there is also a need to develop scheduling strategies for more than two models simultaneously. Overall, the results establish hardware-aware layer reformulation as a viable strategy for enabling fully accelerator-resident multi-model pipelines on embedded heterogeneous platforms. Such reformulations have been proven to improve the performance of medical imaging and diagnosis.


%





\ifCLASSOPTIONcaptionsoff
  \newpage
\fi



%
\bibliographystyle{IEEEtran}
\bibliography{bibtex/bib/IEEEref}

@Article{es1865, 
author="Sali, Safa and Meribout, Anis and Majeed, Ashiyana and Meribout, Mahmoud and Pablo, Juan and Tiwari, Varun and Baobaid, Asma", 
title="Real-Time Object Detection and Associated Hardware Accelerators Targeting Autonomous Vehicles", 
journal="Engineered Science", 
year="2025", 
volume="38", 
pages="1865", 
issn="2576-9898", 
doi="10.30919/es1865", 
url="http://dx.doi.org/10.30919/es1865" 
}

@INPROCEEDINGS{pix2pix,
  author={Isola, Phillip and Zhu, Jun-Yan and Zhou, Tinghui and Efros, Alexei A.},
  booktitle={2017 IEEE Conference on Computer Vision and Pattern Recognition (CVPR)}, 
  title={{Image-to-Image Translation with Conditional Adversarial Networks}}, 
  year={2017},
  volume={},
  number={},
  pages={5967-5976},
  doi={10.1109/CVPR.2017.632},
  key="IEEE"
}

@misc{guide_to_convolution,
      title={{A guide to convolution arithmetic for deep learning}}, 
      author={Vincent Dumoulin and Francesco Visin},
      year={2018},
      eprint={1603.07285},
      archivePrefix={arXiv},
      primaryClass={stat.ML},
      url={https://arxiv.org/abs/1603.07285}, 
      key="IEEE"
}

@misc{deepstream,
  title         = {{DeepStream Documentation}},
  url           = "https://docs.nvidia.com/metropolis/deepstream/dev-guide/text/DS_Overview.html",
  year          = "2025",
  key           = "IEEE",
note = {(Accessed Apr. 17, 2025)}
}

@misc{tensorrt,
  title         = {{TensorRT Documentation}},
  url           = "https://docs.nvidia.com/deeplearning/tensorrt/latest/index.html",
  year          = "2025",
  key           = "IEEE",
note = {(Accessed Apr. 17, 2025)}
}

@misc{nsight,
  title         = {{Nsight Systems Documentation}},
  url           = "https://docs.nvidia.com/nsight-systems/index.html",
  year          = "2025",
  key           = "IEEE",
note = {(Accessed Aug. 25, 2025)}
}

@INPROCEEDINGS{power-better,
  author={Shalavi, Neda and Khoshsirat, Aria and Stellini, Marco and Zanella, Andrea and Rossi, Michele},
  booktitle={2023 IEEE International Conference on Edge Computing and Communications (EDGE)}, 
  title={{Accurate Calibration of Power Measurements from Internal Power Sensors on NVIDIA Jetson Devices}}, 
  year={2023},
  volume={},
  number={},
  pages={166-170},
  doi={10.1109/EDGE60047.2023.00034}}

@misc{onnx-graph,
  title         = {{ONNX GraphSurgeon Documentation}},
  url           = "https://pypi.org/project/onnx-graphsurgeon/",
  year          = "2025",
  key           = "IEEE",
note = {(Accessed Aug. 25, 2025)}
}

@article{Jedi,
author = {Jeong, Eunjin and Kim, Jangryul and Ha, Soonhoi},
title = {{TensorRT-Based Framework and Optimization Methodology for Deep Learning Inference on Jetson Boards}},
year = {2022},
issue_date = {September 2022},
publisher = {Association for Computing Machinery},
address = {New York, NY, USA},
volume = {21},
number = {5},
issn = {1539-9087},
url = {https://doi.org/10.1145/3508391},
doi = {10.1145/3508391},
journal = {ACM Trans. Embed. Comput. Syst.},
month = oct,
articleno = {51},
numpages = {26},
  key           = "IEEE"
}

@inproceedings{HaX-CoNN,
author = {Dagli, Ismet and Belviranli, Mehmet E.},
title = {{Shared Memory-contention-aware Concurrent DNN Execution for Diversely Heterogeneous System-on-Chips}},
year = {2024},
isbn = {9798400704352},
publisher = {Association for Computing Machinery},
address = {New York, NY, USA},
url = {https://doi.org/10.1145/3627535.3638502},
doi = {10.1145/3627535.3638502},
booktitle = {Proceedings of the 29th ACM SIGPLAN Annual Symposium on Principles and Practice of Parallel Programming},
pages = {243–256},
numpages = {14},
location = {Edinburgh, United Kingdom},
series = {PPoPP '24},
  key           = "IEEE"
}

@ARTICLE{pair-unpair,
  title    = {{"Deep {CT} to {MR} Synthesis Using Paired and Unpaired Data"}},
  author   = "Jin, Cheng-Bin and Kim, Hakil and Liu, Mingjie and Jung, Wonmo
              and Joo, Seongsu and Park, Eunsik and Ahn, Young Saem and Han, In
              Ho and Lee, Jae Il and Cui, Xuenan",
  journal  = "Sensors (Basel)",
  volume   =  19,
  number   =  10,
  month    =  may,
  year     =  2019,
  address  = "Switzerland",
  language = "en"
}

@ARTICLE{YOLO_models,
  author={Ragab, Mohammed Gamal and Abdulkadir, Said Jadid and Muneer, Amgad and Alqushaibi, Alawi and Sumiea, Ebrahim Hamid and Qureshi, Rizwan and Al-Selwi, Safwan Mahmood and Alhussian, Hitham},
  journal={IEEE Access}, 
  title={{A Comprehensive Systematic Review of YOLO for Medical Object Detection (2018 to 2023)}}, 
  year={2024},
  volume={12},
  number={},
  pages={57815-57836},
  doi={10.1109/ACCESS.2024.3386826}}

@misc{dataset,
  title         = {{MRI-to-CT-DCNN-TensorFlow}},
  url           = "https://github.com/ChengBinJin/MRI-to-CT-DCNN-TensorFlow",
  year          = "2018",
  key           = "IEEE",
note = {(Accessed Aug. 25, 2025)}
}

@misc{dataset_2,
  title         = {{Paired CT and MRI Dataset for Medical Applications}},
  url           = "https://www.kaggle.com/datasets/29c3607295965ebb030f2d158fec487412d84c82528dd44f8ef956aef35541aa",
  year          = "2025",
  key           = "IEEE",
note = {(Accessed Dec. 01, 2025)}
}

@Article{GAN_vs_Diffusion,
author={Usman Akbar, Muhammad
and Larsson, M{\aa}ns
and Blystad, Ida
and Eklund, Anders},
title={{Brain tumor segmentation using synthetic MR images - A comparison of GANs and diffusion models}},
journal={Scientific Data},
year={2024},
month={Feb},
day={29},
volume={11},
number={1},
pages={259},
issn={2052-4463},
doi={10.1038/s41597-024-03073-x},
url={https://doi.org/10.1038/s41597-024-03073-x}
}

@article{survey_MRI_CT,
title = {{Deep learning based synthesis of MRI, CT and PET: Review and analysis}},
journal = {Medical Image Analysis},
volume = {92},
pages = {103046},
year = {2024},
issn = {1361-8415},
doi = {https://doi.org/10.1016/j.media.2023.103046},
url = {https://www.sciencedirect.com/science/article/pii/S1361841523003067},
author = {Sanuwani Dayarathna and Kh Tohidul Islam and Sergio Uribe and Guang Yang and Munawar Hayat and Zhaolin Chen}
}

@misc{nvdla-doc,
  title         = "{H}ardware {A}rchitectural {S}pecification --- {NVDLA} {D}ocumentation",
  url           = "https://nvdla.org/hw/v1/hwarch.html",
  year          = "2014",
  key           = "IEEE",
note = {(Accessed Apr. 17, 2025)}
}

@misc{xavier-orin-data,
  title         = {{Jetson Download Center}},
  url           = "https://developer.nvidia.com/embedded/downloads",
  year          = "2025",
  key           = "IEEE",
note = {(Accessed Apr. 17, 2025)}
}

@electronic{dla-limit,
  title         = {Working with {DLA} - {DLA} {Supported Layers and Restrictions} },
  url           = "https://docs.nvidia.com/deeplearning/tensorrt/archives/tensorrt-853/developer-guide/index.html#dla_layers",
  year          = "2022",
  key           = "IEEE",
note = {(Accessed Apr. 17, 2025)}
}

@electronic{power,
  title         = {{"Power Management for Jetson Xavier NX Series and Jetson AGX Xavier Series Devices"}},
  url           = "https://docs.nvidia.com/jetson/archives/l4t-archived/l4t-3275/index.html#page/Tegra%20Linux%20Driver%20Package%20Development%20Guide/power_management_jetson_xavier.html",
  year          = "2024",
  key           = "IEEE",
note = {(Accessed Apr. 17, 2025)}
}

@inproceedings{layer_replacement,
  TITLE = {{Energy-efficient use of an embedded heterogeneous SoC for the inference of CNNs}},
  AUTHOR = {Archet, Agathe and Ventroux, Nicolas and Gac, Nicolas and Orieux, Fran{\c c}ois},
  URL = {https://hal.science/hal-04148582},
  BOOKTITLE = {{2023 26th Euromicro Conference on Digital System Design (DSD)}},
  ADDRESS = {Durr{\"e}s, Albania},
  YEAR = {2023},
  MONTH = Sep,
  HAL_ID = {hal-04148582},
  HAL_VERSION = {v1},
}

@Article{accuracy_comparison,
author={Graf, Robert
and Schmitt, Joachim
and Schlaeger, Sarah
and M{\"o}ller, Hendrik Kristian
and Sideri-Lampretsa, Vasiliki
and Sekuboyina, Anjany
and Krieg, Sandro Manuel
and Wiestler, Benedikt
and Menze, Bjoern
and Rueckert, Daniel
and Kirschke, Jan Stefan},
title={{Denoising diffusion-based MRI to CT image translation enables automated spinal segmentation}},
journal={European Radiology Experimental},
year={2023},
month={Nov},
day={14},
volume={7},
number={1},
pages={70},
issn={2509-9280},
doi={10.1186/s41747-023-00385-2},
url={https://doi.org/10.1186/s41747-023-00385-2}
}

@misc{ultralytics,
  title         = "{{Explore Ultralytics YOLOv8}}",
  url           = "https://docs.ultralytics.com/models/yolov8/#overview",
  year          = "2023",
  key           = "IEEE",
note = {(Accessed Aug. 25, 2025)}
}

@misc{yolo-dataset,
  title         = "{{Brain Stroke Detection}}",
  url           = "https://universe.roboflow.com/shreyyy/brain-stroke-detection/",
  year          = "2024",
  key           = "IEEE",
note = {(Accessed Aug. 25, 2025)}
}

@misc{TPU,
  title         = "{{Available TensorFlow Ops}}",
  url           = "https://cloud.google.com/tpu/docs/tensorflow-ops#unavailable_python_apis",
  year          = "2025",
  key           = "IEEE",
note = {(Accessed Sep. 18, 2025)}
}

@misc{Sima,
  title         = "{{ONNX Operators Support List}}",
  url           = "https://docs.sima.ai/pages/model-sdk/supported_onnx_operators.html",
  year          = "2025",
  key           = "IEEE",
note = {(Accessed Sep. 18, 2025)}
}

@ARTICLE{pix2pix-example,
  title    = "{{Magnetic resonance image ({MRI}) synthesis from brain computed
              tomography ({CT}) images based on deep learning methods for
              magnetic resonance ({MR)-guided} radiotherapy}}",
  author   = "Li, Wen and Li, Yafen and Qin, Wenjian and Liang, Xiaokun and Xu,
              Jianyang and Xiong, Jing and Xie, Yaoqin",
  journal  = "Quant Imaging Med Surg",
  volume   =  10,
  number   =  6,
  pages    = "1223--1236",
  month    =  jun,
  year     =  2020,
  address  = "China",
  language = "en"
}

@Article{breast-tumor,
AUTHOR = {Shia, Wei-Chung and Ku, Tien-Hsiung},
TITLE = {{Enhancing Microcalcification Detection in Mammography with YOLO-v8 Performance and Clinical Implications}},
JOURNAL = {Diagnostics},
VOLUME = {14},
YEAR = {2024},
NUMBER = {24},
ARTICLE-NUMBER = {2875},
URL = {https://www.mdpi.com/2075-4418/14/24/2875},
PubMedID = {39767236},
ISSN = {2075-4418},
DOI = {10.3390/diagnostics14242875}
}

@ARTICLE{breast-tumor-2,
  title    = "{{Optimized {YOLOv8} for enhanced breast tumor segmentation in
              ultrasound imaging}}",
  author   = "Mostafa, Ayman Mohamed and Alaerjan, Alaa S and Aldughayfiq,
              Bader and Allahem, Hisham and Mahmoud, Alshimaa Abdelraof and
              Said, Wael and Shabana, Hosameldeen and Ezz, Mohamed",
  journal  = "Discov Oncol",
  volume   =  16,
  number   =  1,
  pages    = "1152",
  month    =  jun,
  year     =  2025,
  address  = "United States",
  language = "en"
}

@article{polyps,
	author = {Hang Dao Viet and Tung Thanh Nguyen and Hoa Ngoc Lam and Binh Phuc Nguyen and Trung Quoc Vu and Hien Minh Nguyen and Vinh Tuan Pho and Hieu Huy Dang and Dinh Viet Sang and Thuy Thi Nguyen},
	title = {{Validation of YOLOv8 algorithm in detecting colon polyps in endoscopy videos}},
	journal = {Journal of Medical Artificial Intelligence},
	volume = {8},
	number = {0},
	year = {2025},
	issn = {2617-2496},	url = {https://jmai.amegroups.org/article/view/10101}
}

@ARTICLE{retinal-imaging,
  title    = "{{Multi-Disease} Detection in Retinal Imaging Based on Ensembling
              Heterogeneous Deep Learning Models}",
  author   = "M{\"u}ller, Dominik and Soto-Rey, I{\~n}aki and Kramer, Frank",
  journal  = "Stud Health Technol Inform",
  volume   =  283,
  pages    = "23--31",
  month    =  sep,
  year     =  2021,
  address  = "Netherlands",
  language = "en"
}

@Article{head-neck-cancer,
AUTHOR = {Turki, Anas and Alshabrawy, Ossama and Woo, Wai Lok},
TITLE = {{Multimodal Deep Learning for Stage Classification of Head and Neck Cancer Using Masked Autoencoders and Vision Transformers with Attention-Based Fusion}},
JOURNAL = {Cancers},
VOLUME = {17},
YEAR = {2025},
NUMBER = {13},
ARTICLE-NUMBER = {2115},
URL = {https://www.mdpi.com/2072-6694/17/13/2115},
PubMedID = {40647415},
ISSN = {2072-6694},
DOI = {10.3390/cancers17132115}
}

@Article{hardware_real_time,
AUTHOR = {Alcaín, Eduardo and Fernández, Pedro R. and Nieto, Rubén and Montemayor, Antonio S. and Vilas, Jaime and Galiana-Bordera, Adrian and Martinez-Girones, Pedro Miguel and Prieto-de-la-Lastra, Carmen and Rodriguez-Vila, Borja and Bonet, Marina and Rodriguez-Sanchez, Cristina and Yahyaoui, Imene and Malpica, Norberto and Borromeo, Susana and Machado, Felipe and Torrado-Carvajal, Angel},
TITLE = {{Hardware Architectures for Real-Time Medical Imaging}},
JOURNAL = {Electronics},
VOLUME = {10},
YEAR = {2021},
NUMBER = {24},
ARTICLE-NUMBER = {3118},
URL = {https://www.mdpi.com/2079-9292/10/24/3118},
ISSN = {2079-9292},
DOI = {10.3390/electronics10243118}
}

@article{computing_acceleration,
author = {Liu, Xing and Dai, Ziheng and Wang, Qi and Li, Zhanpeng},
title = {{Computing Acceleration of Medical Image Processing Based on Multi-Accelerator Heterogeneous Systems}},
year = {2025},
issue_date = {March 2025},
publisher = {Association for Computing Machinery},
address = {New York, NY, USA},
volume = {25},
number = {1},
issn = {1559-6915},
url = {https://doi.org/10.1145/3727257.3727259},
doi = {10.1145/3727257.3727259},
journal = {SIGAPP Appl. Comput. Rev.},
month = apr,
pages = {16–24},
numpages = {9}
}

@ARTICLE{EyerissV2,
  author={Chen, Yu-Hsin and Yang, Tien-Ju and Emer, Joel and Sze, Vivienne},
  journal={IEEE Journal on Emerging and Selected Topics in Circuits and Systems}, 
  title={{Eyeriss v2: A Flexible Accelerator for Emerging Deep Neural Networks on Mobile Devices}}, 
  year={2019},
  volume={9},
  number={2},
  pages={292-308},
  doi={10.1109/JETCAS.2019.2910232}}

@ARTICLE{Epileptic,
  author={Yedurkar, Dhanalekshmi Prasad and Metkar, Shilpa and Al-Turjman, Fadi and Yardi, Nandan and Stephan, Thompson},
  journal={IEEE Transactions on Industrial Informatics}, 
  title={{An IoT-Based Novel Hybrid Seizure Detection Approach for Epileptic Monitoring}}, 
  year={2024},
  volume={20},
  number={2},
  pages={1420-1431},
  doi={10.1109/TII.2023.3274913}}

@ARTICLE{Polyp,
  author={Wang, Dujuan and Wang, Xinwei and Wang, Sutong and Yin, Yunqiang},
  journal={IEEE Transactions on Industrial Informatics}, 
  title={{Explainable Multitask Shapley Explanation Networks for Real-Time Polyp Diagnosis in Videos}}, 
  year={2023},
  volume={19},
  number={6},
  pages={7780-7789},
  doi={10.1109/TII.2022.3208364}}

@article{tpu_vs_gpu,
title = {{Energy efficiency in edge TPU vs. embedded GPU for computer-aided medical imaging segmentation and classification}},
journal = {Engineering Applications of Artificial Intelligence},
volume = {127},
pages = {107298},
year = {2024},
issn = {0952-1976},
doi = {https://doi.org/10.1016/j.engappai.2023.107298},
url = {https://www.sciencedirect.com/science/article/pii/S0952197623014823},
author = {José María {Rodríguez Corral} and Javier Civit-Masot and Francisco Luna-Perejón and Ignacio Díaz-Cano and Arturo Morgado-Estévez and Manuel Domínguez-Morales}
}

@article{Endoscopy,
author = {Xiaoyang Liu  and Parag Karmarkar  and Dirk Voit  and Jens Frahm  and Clifford R. Weiss  and Dara L. Kraitchman  and Paul A. Bottomley },
title = {{Real-Time High-Resolution MRI Endoscopy at up to 10 Frames per Second}},
journal = {BME Frontiers},
volume = {2021},
number = {},
pages = {6185616},
year = {2021},
doi = {10.34133/2021/6185616},
URL = {https://spj.science.org/doi/abs/10.34133/2021/6185616},
eprint = {https://spj.science.org/doi/pdf/10.34133/2021/6185616}}

@Article{Huang2022,
author={Huang, Jiahao
and Ding, Weiping
and Lv, Jun
and Yang, Jingwen
and Dong, Hao
and Del Ser, Javier
and Xia, Jun
and Ren, Tiaojuan
and Wong, Stephen T.
and Yang, Guang},
title={{Edge-enhanced dual discriminator generative adversarial network for fast MRI with parallel imaging using multi-view information}},
journal={Applied Intelligence},
year={2022},
month={Oct},
day={01},
volume={52},
number={13},
pages={14693-14710},
issn={1573-7497},
doi={10.1007/s10489-021-03092-w},
url={https://doi.org/10.1007/s10489-021-03092-w}
}

@INPROCEEDINGS{edgeYOLO,
  author={Liu, Shihan and Zha, Junlin and Sun, Jian and Li, Zhuo and Wang, Gang},
  booktitle={2023 42nd Chinese Control Conference (CCC)}, 
  title={EdgeYOLO: An Edge-Real-Time Object Detector}, 
  year={2023},
  volume={},
  number={},
  pages={7507-7512},
  doi={10.23919/CCC58697.2023.10239786},
  ISSN={1934-1768},
  month={July},}

@misc{Libre,
  author = {Libre Computer},
  title = {{Alta AML-A311D-CC}},
  year = {2024},
  url = {https://libre.computer/products/aml-a311d-cc/},
  key           = "IEEE",
note = {(Accessed Jul. 09, 2025)}
}

@Article{MTPerception_AD,
AUTHOR = {Wang, Hai and Li, Jiayi and Dong, Haoran},
TITLE = {A Review of Vision-Based Multi-Task Perception Research Methods for Autonomous Vehicles},
JOURNAL = {Sensors},
VOLUME = {25},
YEAR = {2025},
NUMBER = {8},
ARTICLE-NUMBER = {2611},
URL = {https://www.mdpi.com/1424-8220/25/8/2611},
PubMedID = {40285295},
ISSN = {1424-8220},
DOI = {10.3390/s25082611}
}

@ARTICLE{FlexNPU,
    AUTHOR={Raha, Arnab  and Mathaikutty, Deepak A.  and Kundu, Shamik  and Ghosh, Soumendu K. },
    TITLE={FlexNPU: a dataflow-aware flexible deep learning accelerator for energy-efficient edge devices},
    JOURNAL={Frontiers in High Performance Computing},
    VOLUME={Volume 3 - 2025},
    YEAR={2025},
    URL={https://www.frontiersin.org/journals/high-performance-computing/articles/10.3389/fhpcp.2025.1570210},
    DOI={10.3389/fhpcp.2025.1570210},
    ISSN={2813-7337}}

@misc{cyclegan,
      title={Unpaired Image-to-Image Translation using Cycle-Consistent Adversarial Networks}, 
      author={Jun-Yan Zhu and Taesung Park and Phillip Isola and Alexei A. Efros},
      year={2020},
      eprint={1703.10593},
      archivePrefix={arXiv},
      primaryClass={cs.CV},
      url={https://arxiv.org/abs/1703.10593}, 
}

@InProceedings{LP,
  author =	{Pujol, Roger and Tabani, Hamid and Kosmidis, Leonidas and Mezzetti, Enrico and Abella, Jaume and Cazorla, Francisco J.},
  title =	{{Generating and Exploiting Deep Learning Variants to Increase Heterogeneous Resource Utilization in the NVIDIA Xavier}},
  booktitle =	{31st Euromicro Conference on Real-Time Systems (ECRTS 2019)},
  pages =	{23:1--23:23},
  series =	{Leibniz International Proceedings in Informatics (LIPIcs)},
  ISBN =	{978-3-95977-110-8},
  ISSN =	{1868-8969},
  year =	{2019},
  volume =	{133},
  editor =	{Quinton, Sophie},
  publisher =	{Schloss Dagstuhl -- Leibniz-Zentrum f{\"u}r Informatik},
  address =	{Dagstuhl, Germany},
  URL =		{https://drops.dagstuhl.de/entities/document/10.4230/LIPIcs.ECRTS.2019.23},
  URN =		{urn:nbn:de:0030-drops-107608},
  doi =		{10.4230/LIPIcs.ECRTS.2019.23},
  key           = "IEEE"
}

@ARTICLE{my_review,
  author={Majeed, Ashiyana Abdul and Meribout, Mahmoud and Sali, Safa Mohammed},
  journal={IEEE Transactions on Industrial Informatics}, 
  title={Scheduling Techniques of AI Models on Modern Heterogeneous Edge GPU—A Critical Review}, 
  year={2026},
  volume={},
  number={},
  pages={1-12},
  doi={10.1109/TII.2025.3642530}}

@ARTICLE{11316238,
  author={Baobaid, Asma and Meribout, Mahmoud},
  journal={IEEE Transactions on Emerging Topics in Computing}, 
  title={Leveraging Simultaneous Usage of Edge GPU Hardware Engines for Video Face Detection and Recognition}, 
  year={2025},
  volume={},
  number={},
  pages={1-11},
  doi={10.1109/TETC.2025.3644672}}
%








\end{document}